\newcommand{\bra}[1]{\langle #1|} \newcommand{\ket}[1]{|#1\rangle}

\documentclass[a4paper,aps, prb,amsmath,amssymb, superscriptaddress, showpacs, twocolumn, floatfix]{revtex4-1}
\usepackage{graphicx, braket, mathrsfs, nicefrac, epsfig, multirow, ulem, verbatim, subfigure}

\begin{document}
\title{Coulomb Blockade due to Quantum Phase-Slips Illustrated with Devices}

\author{A.~M.~Hriscu}
\affiliation{Kavli Institute of Nanoscience, Delft University of
Technology,\\
PO Box 5046, 2600 GA Delft, The Netherlands}
\author{Yu.~V.~Nazarov}
\affiliation{Kavli Institute of Nanoscience, Delft University of
Technology,\\
PO Box 5046, 2600 GA Delft, The Netherlands}

\begin{abstract}
In order to illustrate the emergence of Coulomb blockade from coherent quantum phase-slip processes in thin superconducting wires, we propose and theoretically investigate two elementary setups, or "devices". The setups are derived from Cooper-pair box and Cooper-pair transistor, so we refer to them as QPS-box and QPS-transistor, respectively.

We demonstrate that the devices exhibit sensitivity to a charge induced by a gate electrode, this being the main signature of Coulomb blockade. Experimental realization of these devices will unambiguously prove the Coulomb blockade as an effect of coherence of phase-slip processes. We analyze the emergence of discrete charging in the limit of strong phase-slips. We have found and investigated six distinct regimes that are realized depending on the relation between three characteristic energy scales: inductive and charging energy, and phase-slip amplitude. For completeness, we include a brief discussion of dual Josephson-junction devices. 
\end{abstract}

\maketitle

\section{Introduction}

The interplay between superconductivity and Coulomb interactions in mesoscopic superconducting circuits is a subject of active research for more than 20 years \cite{AverinLikharev, LesHouses, Saclay, Haviland1989,Fazio2001}. 
A generic device that has been widely studied and used is the Cooper-pair box (CPB) \cite{Saclay, Buttiker1987, Bouchiat1998}. It consists of a superconducting island to store discrete charges that is connected to a bulk electrode by means of a tunnel  Josephson junction. This enables coherent transfer of Cooper pairs between the island and electrode. A gate electrode capacitively coupled to the island induces a continuous charge $q$. The energies of quantum states of the CPB are periodic functions of this induced charge. This {\it charge sensitivity} reveals the charging states of the device, those  with a well defined number of excess Cooper pairs, and opens up the possibility to create and control their quantum superpositions. For instance, the CPB can be operated near the point where two charging states are approximately degenerate forming a qubit basis. The Josephson tunneling lifts the degeneracy between these charging states and enables their superpositions.\cite{Nakamura1997}

CPB  and similar devices based on Coulomb blockade, like flux qubits \cite{MooijOrlando}, have been used to realize viable qubit schemes. Among other achievements, coherent control of quantum states \cite{Nakamura1999}, Rabi oscillations \cite{Nakamura2001, Martinis2002}, successful DC readout \cite{Vion2002, Duty2004} and RF coupling of multiple junctions have been demonstrated.

The studies of phase-slip processes \cite{review} in superconducting wires have a long and spectacular history. 
During a phase-slip process, the superconducting order parameter passes zero at a certain moment of time and at a certain position in the wire. The phase difference between the wire ends changes by $2 \pi$. In accordance with Josephson relation, this gives rise to a voltage pulse across the wire. It has been established more than 40 years ago \cite{Little} that the residual resistance of the wires below the superconducting transition is due to thermally activated phase-slips.\cite{LAMH} 

At sufficiently low temperatures, the quantum fluctuations should supersede the thermally-activated ones. The manifestations of these quantum phase-slips in ultra-thin resistive wires have been actively investigated for the last 10 years \cite{Bezryadin2000,Lau2001, Altomare2006, Bollinger2008}. It has been suggested that the switching from superconducting to normal state in current-driven wires is caused by individual quantum phase-slips events \cite{Bollinger2008,Sahu2009}. These experiments implemented resistive measurements where the quantum coherence between individual phase-slip events seems to be destroyed by accompanying dissipation. If such coherence is preserved, the manifestations of phase-slips is qualitatively different: they change the characteristics of the ground and excited quantum states.\cite{Buchler2004} It has been proposed \cite{MooijHarmans} that the coherent phase-slips should be observed and studied in non-driven devices of qubit type. 
Recently, the effect of coherent quantum phase-slips has been observed in Josephson chains that are in many respects similar to superconducting wires\cite{Pop2010}.

\begin{figure}[!b] 
\includegraphics[width=0.65\columnwidth]{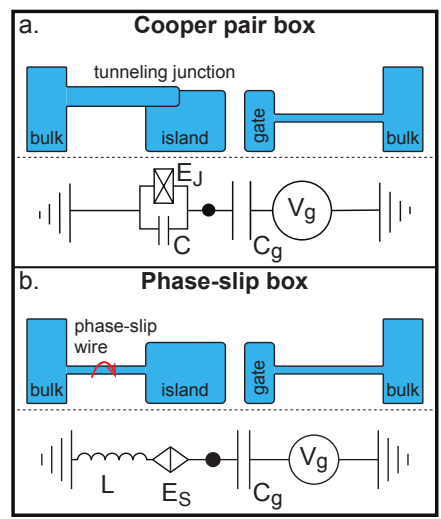}
\caption{a. Cooper-pair box(CPB): generic Coulomb-blockade superconducting device. b. In quantum phase-slip box (QPS-box), the tunnel junction is replaced with  a superconducting wire.}
\label{Fig1}
\end{figure}

It is clear that sufficiently well-developed coherent phase-slips should eventually lead to Coulomb blockade in the wire. This statement is fascinatingly counterintuitive: the superconducting wire is one of the best conductors possible, while Coulomb blockade requires isolation. However, the necessity of Coulomb blockade and accompanying isolation in this regime directly follows from basic arguments that involve duality between the charge and phase \cite{Mooij2006}. In a nutshell, the argumentation is as follows. The coherence of Cooper-pair tunneling events leads to a zero-voltage states at currents below a critical value. Interchanging charge and phase, current and voltage, we prove that the coherence of the phase-slip events should lead to a zero-current state below a critical voltage. This is essentially Coulomb blockade and isolation.
Indeed, there are experimental data that indicate a possible crossover to insulating behavior in ultra-thin wires.\cite{Lau2001,Bollinger2008} However, the exact conditions and mechanism of this crossover are still subject to debate.\cite{debat}

The main motivation of this article is to facilitate an unambiguous experimental proof of Coulomb blockade due to phase-slips. 
In this context, we note that Coulomb blockade is usually accompanied by periodic charge sensitivity. The observation of charge sensitivity in setups where the tunnel barriers are replaced by uninterrupted superconducting wires would constitute the proof required. The experimental attention to charge sensitivity is presently insufficient, though a very recent communication\cite{DevoretNew} reports indirect observation of charge sensitivity by its effect on the decoherence in fluxonium\cite{fluxonium} qubit.
We have also recently learned of unpublished results of the authors of Ref.
\cite{Pop2010} that demonstrate the gate-voltage effect.
In this article, we analyze the problem at elementary level by introducing simple phase-slip systems (we call them devices)
where this charge sensitivity can be observed, and discuss the conditions for this to happen.

\begin{figure}[!t]
\centering
\includegraphics[width=0.8\columnwidth]{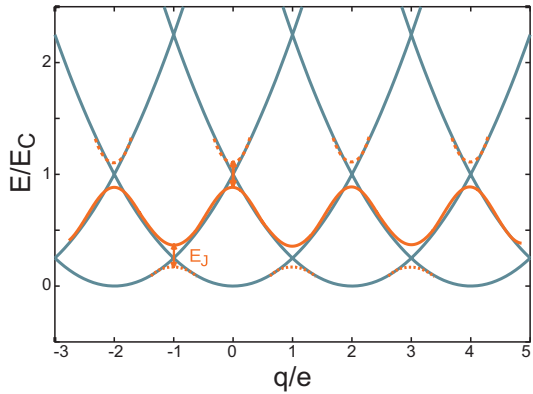}
\caption{Energy spectrum of CPB versus induced charge $q$ in the limit of large charging energy. The discrete charging states give rise to a standard $2e$-periodic pattern of crossing parabolas. The spectrum of phase-slip devices considered in the article follows the same pattern in the limit of large phase-slip amplitudes.}
\label{fig:CPB-parabolas}
\end{figure}

\begin{figure}[!b] 
\includegraphics[width=0.65\columnwidth]{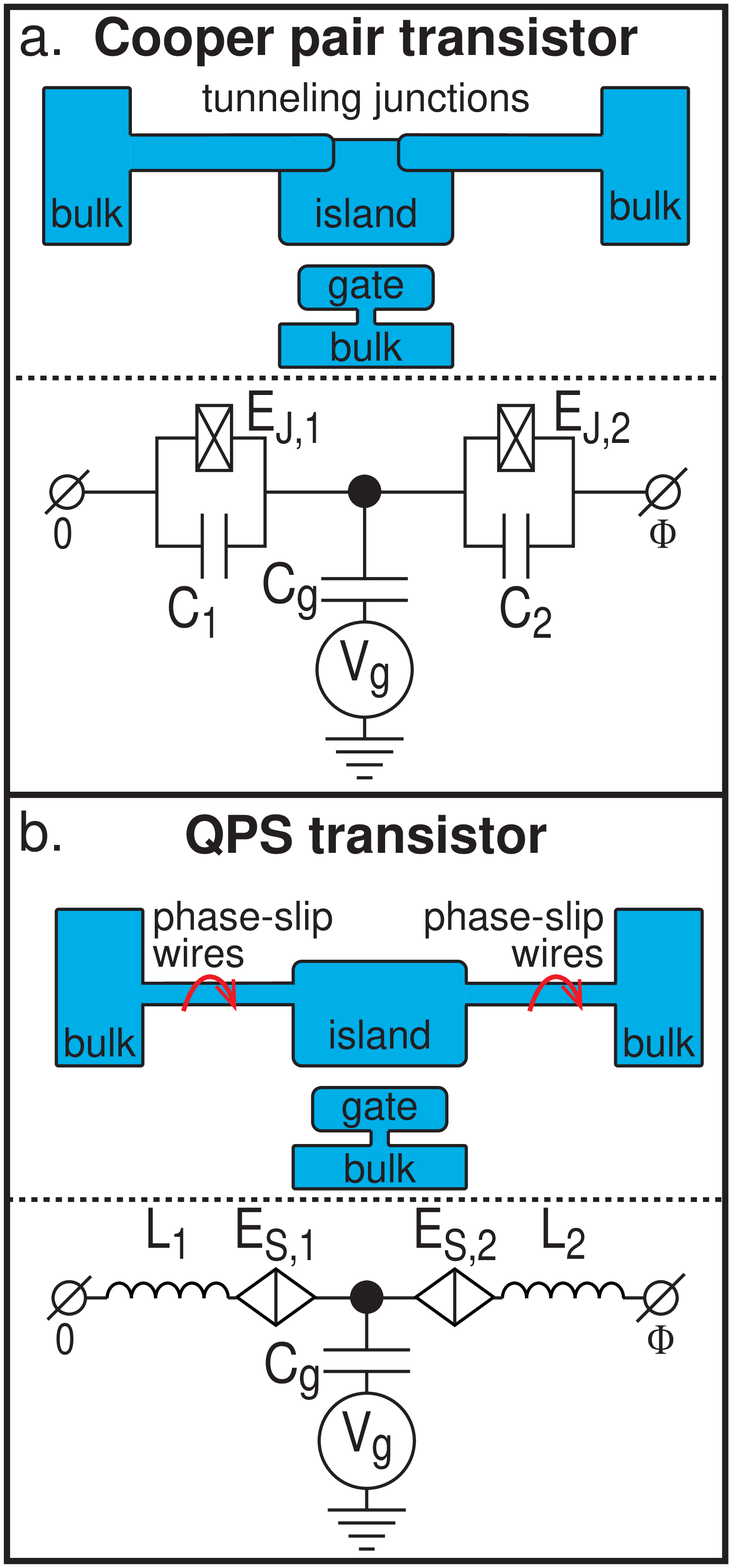}
\caption{a. Cooper-pair transistor (CPT). The dependence of its energy on external phase $\Phi$ (flux sensitivity) gives rise to the supercurrent that can be modulated by changing $q=C_g V_g$. b. Quantum phase-slip transistor.}
\label{Fig3}
\end{figure}

The devices are obtained from the generic Cooper-pair box (CPB) and Cooper-pair box with two leads (Cooper-pair transistor, CPT)
by replacing the tunnel junctions with the superconducting wires subject to phase-slips. We ascribe all the capacitance to the island, thus disregarding geometric capacitance of the wires. We also disregard quasiparticle effects assuming that the superconducting gap by far exceeds the temperature. These assumptions allows us to account for the phase-slips in the framework of a zero-dimensional phenomenological model of phase-slip junction\cite{Mooij2006} (we discuss the relation of this model and microscopic theory of phase-slips in the Appendix). We call the devices QPS-box and QPS-transistor. 

 Let us note that for the devices proposed, the Coulomb blockade and associated charging states do not occur {\it in} the wire. It would be wrong
to assume that the wire is broken into a chain of weakly connected Coulomb islands. Indeed, the assumption of vanishing self-capacitance in fact forbids any charge accumulation in the wire. Rather, the collective state developed in the wire as a whole provides the isolation between the wire ends. Qualitatively, the wire as a whole acts as a single tunnel barrier.

We study the energy levels in these devices, with emphasis on the ground state.
The operational limit of the devices is set by the value of the effective impedance $\gamma = (E_L/E_C )^{1/4}$: the ratio of the inductive and capacitive energy. In the limit of large impedance, the phase fluctuations are dominant, while for small impedance regime the charge fluctuations become relevant. The values of $L$ and $C$ are set by fabrication and they have a wide range of possible values. In order to make qualitative and quantitative predictions about the devices we compare the corresponding inductive and charging energies with the phase-slip amplitude.

We demonstrate that these energies exhibit charge sensitivity (for the boxes)
and combined flux/charge sensitivity (for the transistors).
Moreover, in the limit of large phase-slip amplitude, the phase-slip devices can be directly mapped on their tunnel-junction counterparts, and exhibit the standard pattern of "crossing parabolas": charge-sensitive discrete charging states.

For completeness, let us make a remark about the effect of the random offset charges on our devices. Without phase-slips the proposed devices are just linear and not affected by any charge. However, the charge sensitivity brought about by the phase-slips implies that the devices are affected not only by the gate voltage but also by the random offset charges most likely present in the substrate. In general, it is a task of the fabrication technology to minimize the effect of offset charges. We expect, that the effect of the offset charges on the devices under consideration would be the same as that on any other Coulomb blockade system like SET transistor, quantum dot, superconducting qubit etc. At sufficiently low substrate temperature the random offset charges remain the same and can be compensated for by a shift in the gate voltage. If the devices are realized with metal wires (rather than with Josephson junction chains) there is a chance that the offset charges are efficiently screened by the metal leads. Therefore, such realization may be better than that involving tunnel junctions where the offset charges are present in the insulating layers of the junctions.

The paper is organized as follows. Section \ref{sec:description} describes in detail the devices under consideration and establishes their Hamiltonians. Section \ref{sec:QPS-box} is devoted to QPS-box. We consider first the phase-slips perturbatively, then analyze the crossover to Coulomb blockade in the regimes of large, small and intermediate impedance. Section \ref{sec:QPS-transistor} details the QPS-transistor with emphasis on flux sensitivity specific for this device. In Section \ref{sec:dual} we discuss the Josephson-based devices that are dual to QPS-box and QPS-transistor. We conclude in Section \ref{sec:conclusions}. 

\section{Description of the devices}\label{sec:description}
Let us recall two generic devices that exemplify the manifestation of Coulomb blockade in superconducting circuits. They are made by connecting 
a superconducting island with either one or two superconducting leads. An isolated island supports discrete charges. The important part of the setup is the gate electrode 
that is not electrically connected to the island but, by means of capacitive coupling, induces charge $q$ on the island. We will refer to these two devices as to Cooper-pair box (CPB, Fig. \ref{Fig1}.a.) and Cooper-pair transistor (CPT). The latter term is less conventional: we use it because the supercurrent through the device does depend on the gate voltage, this being a transistor effect. Besides, the setup reminds much that of normal-metal Single-Electron Tunneling Transistor (SET)\cite{FultonDolan}. 

Our idea of introducing the phase-slip devices is to replace the tunnel junctions in above setups with thin superconducting wires.
In this way, we come to the setups of QPS-box and QPS-transistor.
Healthy reasoning is that the wires would short-circuit superconducting island to the lead or leads: unlike the tunnel junctions, the wires are not expected to provide isolation required for Coulomb blockade phenomena. In this case, the charge induced to the island should have no physical effect. 
This is indeed true if coherent phase-slips in the wires are disregarded. In this case, both devices are just linear electric circuits.

It is the main goal of the present manuscript to show in detail that the coherent phase-slips induce charge sensitivity in both devices. Moreover, for sufficiently large phase-slip amplitudes, the phase-slip devices are eventually identical to CPB and CPT, respectively, and exhibit almost pure charging states like shown in Fig. \ref{fig:CPB-parabolas}. For QPS-transistor, the charge sensitivity is combined with the flux sensitivity, so one can observe a transistor effect.

All devices are simple to describe quantum-mechanically: this involves a Hamiltonian with several degrees of freedom only. In the rest of this Section, we give these Hamiltonians as the basis for further consideration.

\subsection{CPB versus QPS-box}

The Hamiltonian of the CPB consists of charging and Josephson terms, 
\begin{equation}
\hat{H}_{\rm{CPB}} = E_C \left(\hat{Q}- \frac{q}{2e} \right)^2 -E_J \cos (\hat{\phi}),
\label{eq:H_CPB}
\end{equation}
where $E_C = \frac{2e^2}{(C+C_g)}$ is the charging energy involving the total capacitance  of the island. The Josephson energy can be expressed in terms of   the conductance of the tunnel junction $G$, $E_J = \frac{G}{G_Q}\frac{\Delta}{4}$, where $G_Q$ is the conductance quantum,  and $\Delta$ the superconducting gap. The operator $\hat{Q}$ is the charge stored on the island measured in units of Cooper-pair charge $2e$, $\phi$ is the operator of the superconducting phase difference between the island and the lead. These two operators satisfy the canonical commutation relations $[\hat{Q},\hat{\phi}]=-i$. 

This Hamiltonian is undefined unless we specify the space in which the variables $Q,\phi$ are defined. For CPB, the phase is defined in the interval $(-\pi,\pi)$ so that the wave function is periodic in $\phi$, $\psi(\phi)=\psi(\phi+2\pi)$. This assumption makes the charge variable discrete, $Q=N$, $N$ being integer numbers spanning a countable set of states $|N\rangle$. The Hamiltonian in charge representation then reads
\begin{align}
\label{eq:H_CPB_charge}
\hat{H}_{\rm{CPB}} &= \sum_N E_C \Big(N- \frac{q}{2e}\Big)^2 \ket{N}\bra{N} \notag \\
& \ \ \ \ \ - \frac{ E_J}{2} \left(\ket{N}\bra{N+1}+ h.c.\right).
\end{align}

The QPS-box includes a superconducting wire. If we restrict ourselves to the energies lower than the inverse time of
propagation of electric excitations along the wire, we can neglect the details of spatial distribution of the order parameter in the wire and characterize its quantum state by a single quantum variable $\phi$ that correspond to the phase drop across the wire.
Neglecting the details of the spatial distribution implies that we the disregard geometric capacitance of the wire. More precisely, we ascribe all the capacitance to the island. We stress that in distinction from CPB the phase variable is extended being defined in the interval $(-\infty,\infty)$.
One can regard this variable as consisting from a compact phase of the island and discrete number of phase windings $p$ along the superconducting wire. Correspondingly, the charge is a continuous variable, and Coulomb blockade is not evident. Without phase-slips, the wire can be regarded as a linear inductor of inductance $L$. Together with the island capacitance, this gives an LC oscillator. The phase-slips change the winding number $p$ by $\pm 1$. Since we neglect the details of spatial distribution, it does not matter where in the wire a phase-slip would occur. The tunneling between different $p$ can be thus described  by a single amplitude.\cite{Mooij2006}
We present more microscopic details in the Appendix.

These assumptions define the Hamiltonian of the QPS-box. It consists of a part that describes a linear circuit, in our case, an LC-oscillator, and phase-slips
\begin{equation}\label{eq:H_PSB}
H_{\rm{PSB}} = E_C \left(\hat{Q}-\frac{q}{2e}\right)^2 + \frac{E_L}{4} \hat{\phi}^2 + H_S, 
\end{equation}
where
\begin{equation*}
E_C = \frac{2 e^2}{C} ; \quad E_L = \frac{2}{L} \left (\frac{\Phi_0}{2 \pi}\right)^2,
\end{equation*}
are the charging energy corresponding to charging the capacitor $C$ and respectively the inductive energy corresponding to the inductor $L$, $\Phi_0$ being the flux quantum.

The action of the phase-slip Hamiltonian adds $\pm 1$ to the winding number, that is, shifts the wavefunction in phase variable by $\pm 2\pi$:
\begin{equation} \label{eq:1PS_hamilt}
H_S \ \psi(\phi) = -E_S \ \psi(\phi+2\pi) - E_S\ \psi(\phi-2\pi),
\end{equation}
$E_S$ being the phase-slip amplitude.
In the charge representation this term takes a simple form of a cos-potential:

\begin{equation}\label{eq:Hamilt}
H_S = -2 E_S \cos (2\pi \hat{Q}).
\end{equation}

The charge sensitivity is entirely due to the phase-slip term: one can make this explicit by shifting the charge variable by $q/2e$.
The induced charge $q$ disappears from the oscillator term while the phase-slip amplitudes acquire phase factors:

\begin{equation} \label{eq:1PS_hamilt-a}
H_S \psi(\phi) = -E_S e^{-i\pi q/e} \psi(\phi+2\pi) - E_S  e^{i\pi q/e} \psi(\phi-2\pi).
\end{equation}

One can say that the induced charge affects the interference of phase-slips of two opposite directions.
It is a matter of choice whether Eq. \ref{eq:1PS_hamilt} of Eq. \ref{eq:1PS_hamilt-a} defines the phase-slip operator since the difference is the shift in charge space.
Depending on the choice, the charging energy reads
either $E_C (\hat{Q} -q/2e)$ or $E_C \hat{Q}$.

We notice that the QPS-box without phase-slips ($E_S=0$) is a $LC$ oscillator with the frequency 
$$
\hbar \omega_0 = \frac{\hbar}{\sqrt{LC}} =  \hbar \sqrt{E_L E_C}
$$
and the spectrum
$$
E_n= \hbar \omega_0 \left( n+ \frac{1}{2} \right),
$$
$n$ being number of oscillator quanta.
The oscillator is characterized by an important parameter $\gamma$ defined as 

\begin{equation*}
\gamma = \sqrt{\frac{\hbar}{2 e^2} \sqrt{\frac{C}{L}}}= \sqrt{\frac{1}{2 \pi G_Q Z}} 
\end{equation*}
that measures the effective impedance $Z$ of the oscillator in quantum units. This parameter can be readily expressed through the ratio of the charging and inductive energies: $\gamma^2 =\sqrt{\frac{E_L}{E_C}}$. In terms of the creation-annihilation operators $b,b^{\dagger}$ of the oscillator the phase-slip Hamiltonian reads $H_S =-E_S (e^{\pi \gamma (b^{\dagger}-b)} + h.c.) $.

\subsection{CPT versus QPS-transistor}
Physically, the CPT is a device that is more functional than the CPB. It can conduct the superconducting current that is affected by the induced charge, this presents a convenient way to detect the charge sensitivity. However, at Hamiltonian level the devices are described by the same Eq. \ref{eq:H_CPB_charge}. The only difference is that the Cooper-pair tunneling to/from the island can proceed through two junctions, both adding to the tunneling amplitude,
$$
E_J = E_{J,1} + E_{J,2} e^{i\Phi}
$$ 
The superconducting phase difference between the leads $\Phi$ affects the interference of the two amplitudes.  We note that a practical way to realize such phase bias is to embed the device into a superconducting loop penetrated by magnetic flux so that we interchangeably refer to this parameter as to phase or flux. For instance, we talk about flux sensitivity while referring to the dependence of energy levels on $\Phi$.

In contrast to this, the Hamiltonian description of the QPS-transistor (Fig. \ref{Fig3}.b) is more complex than that of the QPS-box. 
The point is that there are two wires in the device, each with its own winding number. This gives rise to two extended phases $\phi_{1,2}$. However, there is a constrain on these phases. They should sum up to the overall phase drop over the device that is fixed externally. It is convenient to restrict the external phase to the interval $(-\pi,\pi)$. In this case, the constrain reads

\begin{equation}\label{eq:phs-cond}
\phi_1 + \phi_2 = \Phi + 2 \pi p,
\end{equation}

integer $p$ being the total winding number in both wires.

Let us introduce two continuous charge variables $Q_{1,2}$ that are canonically conjugated to these phases. With this, the Hamiltonian of the QPS-transistor reads
\begin{align}
H_{\rm{PST}} &= \frac{E_{L,1}}{4} \hat{\phi}_1^2 + \frac{E_{L,2}}{4} \hat{\phi}_2^2 \notag \\
&  + E_{C} (\hat{Q}_2 -\hat{Q}_1 - \frac{q}{2e})^2  \notag\\
&  + H_{S,1} + H_{S,2} \label{eq:H0-PSET} ,
\end{align}
where the phase-slip Hamiltonians are
\begin{equation*}
H_{S, 1(2)} = - 2 E_{S,1 (2)} \cos{(2 \pi \hat{Q}_{1 (2)})}.
\end{equation*}

We recognize the inductive terms, proportional to the squares of the phase drops $\phi_1, \phi_2$ across each inductor. The total charge accumulated in the island is the difference of the charges $Q_2,Q_1$ passing the wires, this gives the form of the charging energy. The last two terms present the phase-slips that shift the corresponding phase drops by $\pm 2 \pi$. The Hamiltonian can be written in several equivalent representations. In charge representation, the constrain (\ref{eq:phs-cond}) implies the the following periodicity condition on the wave function,
$$
\psi(Q_1,Q_2) = \psi(Q_1+1,Q_2+1)
$$ 
while the Hamiltonian reads

\begin{align}
H_{\rm{PST}} &= \frac{E_{L,1}}{4} \left(-i\frac{\partial}{\partial Q_1}\right)^2 + \frac{E_{L,2}}{4}\left(-i\frac{\partial}{\partial Q_2} + \Phi\right)^2\notag \\
&  + E_{C} \left(Q_2 -Q_1 - \frac{q}{2e}\right)^2  \notag\\
&  - 2 E_{S,1}\cos(2\pi Q_1) - 2 E_{S,2}\cos(2\pi Q_2)\label{eq:H0-PSET-charge}.
\end{align}

However, the properties of the QPS-transistor at $E_{S,1}, E_{S,2}$ are easier to understand in phase representation.
In this case, the wavefunction is defined at a series of lines numbered by $p$ and parametrized with continuous $\phi_1$.
At each line, the wavefunctions are that of a harmonic oscillator with the frequency
$$
\hbar \omega_0 = \sqrt{E_C (E_{L,1}+E_{L,2})}.
$$
The ground state energies of these oscillators are different on different lines. They correspond to minimum inductive energy at
a given total winding number $p$,
$$
E_p=\frac{E_{L,1} \cdot E_{L,2}}{4(E_{L,1}+E_{L,2})} (2 \pi p + \Phi)^2. 
$$
Therefore the spectrum reads
\begin{equation}\label{eq:E0-PSET}
E_{n,p}= \hbar \omega \Big(n+ \frac{1}{2} \Big) + E_p.
\end{equation}
It is important to recognize that the oscillators at different p are shifted with respect to each other,
their equilibrium positions being given by 
$$
\phi_1^{(0)}(p) = \frac{E_{L,2}}{E_{L,1}+E_{L,2}} \cdot (2 \pi p + \Phi).
$$

In this representation, the phase-slip Hamiltonians either increase or decrease the total winding number, while
$H_{S,1}$ also shifts $\phi_1$ by $\pm 2\pi$,

\begin{align}
H_{S,1} \psi(\phi_1,p) &= -2 E_{S,1} \Psi(\phi_1+ 2 \pi,p+1) \notag\\
&\ \ \ -2 E_{S,1}^{*} \Psi(\phi_1 -2 \pi,p-1)\label{eq:HS1},\\
H_{S,2} \psi(\phi_1,p) &= -2 E_{S,2} \psi(\phi_1,p+1) -2 E_{S,2}^{*} \psi(\phi_1,p-1)\label{eq:HS2}
\end{align}

We define the parameter $\gamma$ that measures the effective impedance of the oscillator $\gamma^2 = \sqrt{(E_{L,1}+E_{L,2})/E_C}$.

In contrast to the QPS-box, the QPS-transistor exhibits both flux and charge sensitivity that makes it potentially useful for measurements.

\section{QPS-box} \label{sec:QPS-box}

In this Section, we study the ground state and low-energy states of QPS-Cooper pair box, with emphasis on the charge sensitivity of their energies.  The charge sensitivity appears already in the limit of $E_S \to 0$ as a first-order perturbation correction to the oscillator levels. Upon increasing $E_S$, the charge sensitivity increases and  eventually the low-lying states follow the standard Coulomb-blockade pattern. The crossover to the Coulomb blockade follows different scenarios depending on the effective impedance of the oscillator $1/\gamma^2$. We discuss the limits of large and small impedance and present numerical illustrations for these limits as well as for the case of intermediate impedance.

\subsection{First-order corrections}
In the limit of $E_S\ll E_C, E_L$ it should be possible to treat the phase-slip term $H_{S}$ as a perturbation. 
The unperturbed system at $E_S=0$ is nothing but an $LC$ oscillator, and the first order correction is given by the diagonal matrix element of $H_S$, $E^{(1)}_n=\langle n|\hat{H}_S|n \rangle$, with respect to these states.
The correction gives the charge-sensitive part of the energy and reads
\begin{equation}\label{eq:PSB-corr}
E^{(1)}_n =- 2 E_S \cos \left(\frac{\pi q}{e}\right) \exp(-\pi^2 \gamma^2/2)\ _1F_1[-n,1,\gamma^2 \pi^2].
\end{equation}
Here $_1 F_1$ stands for the confluent hypergeometric function of the first kind. 

\begin{figure}[!t]
\centering
\includegraphics[width=0.8 \columnwidth]{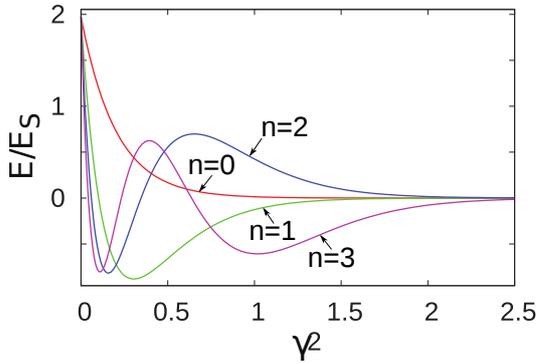}
\caption{The magnitude of the charge-sensitive first-order correction (Eq. \ref{eq:PSB-corr}) to the first four energy levels $n=0, \ldots 4,$ versus $\gamma^2$.}
\label{fig:encorr} 
\end{figure}

This expression is valid at any value of the effective impedance. In Fig. \ref{fig:encorr} we plot this correction for the first four eigenenergies versus $\gamma^2$.
For the ground state $n=0$, the correction reads
\begin{equation} \label{eq:PSB-corr-ground}
E^{(1)}_{0}=  -2 E_S \cos \big(\frac{\pi q}{e} \big) \exp \left(-\frac{\pi^2}{2} \sqrt{\frac{E_L}{E_C}}\right).
\end{equation}
we we use $\gamma^2 =\sqrt{\frac{E_L}{E_C}}$.

The first-order corrections have a characteristic $\cos(\pi q/ e )$ dependence on the induced charge. Despite their relatively small magnitude, they can be revealed by driving the oscillator with a.c.  gate voltage at a frequency close to $\omega_0$.\cite{Hriscu2009}
The first-order corrections are exponentially suppressed in the limit of small effective impedance, $\gamma \gg 1$.
To see that, let us estimate a typical spread of the wave function of the ground state in phase space, that is, quantum fluctuation of phase. For the case of small effective impedance, this fluctuation is small, $\simeq 1/\gamma$. The first-order correction due to phase-slips is given by the overlap of this ground state wave function with its copy shifted by $2\pi$. The small spread leads to exponentially small overlap and thus to the exponentially small correction.

Comparing these first-order corrections with oscillator energy differences $\hbar \omega_0$ suggests that the crossover to well-developed Coulomb blockade takes place at  $E_S \simeq \hbar\omega_0$ for $\gamma \lesssim 1$ and at exponentially large $E_S$ for $\gamma \gg 1$. This estimation is too simplistic: we will consider below the cases of small and large effective impedance to show where and how the crossover actually happens.

\subsection{Small impedance regime}
\label{sec:es_GAM}
In this case, $E_L \gg E_C$ and thus $\gamma \gg 1$. To understand the specifics of the regime, 
let us first set the charging energy $E_C$ to zero. The Hamiltonian in charge representation reads 
\begin{equation*}
H = -\frac{E_L}{4} \frac{\partial^2}{\partial Q^2} -2 E_S \cos \left(2 \pi\left(Q-\frac{q}{2e}\right)\right),
\end{equation*}
is translational invariant in $Q$-space, and is equivalent to a Hamiltonian of a quantum particle a periodic potential. Owing to translational invariance, the eigenenergies do not depend on $q$ showing no charge sensitivity.
The eigenfunctions are Bloch states labeled by $\chi$, $\psi(Q+1)=e^{i\chi}\psi(Q)$. We will call the parameter $\chi$ the quasiphase, since it is similar to the variable $\phi$ and plays the role of quasi-momentum for common Bloch states in periodic solids.

\begin{figure}[!t]
\centering
\includegraphics[width=\columnwidth]{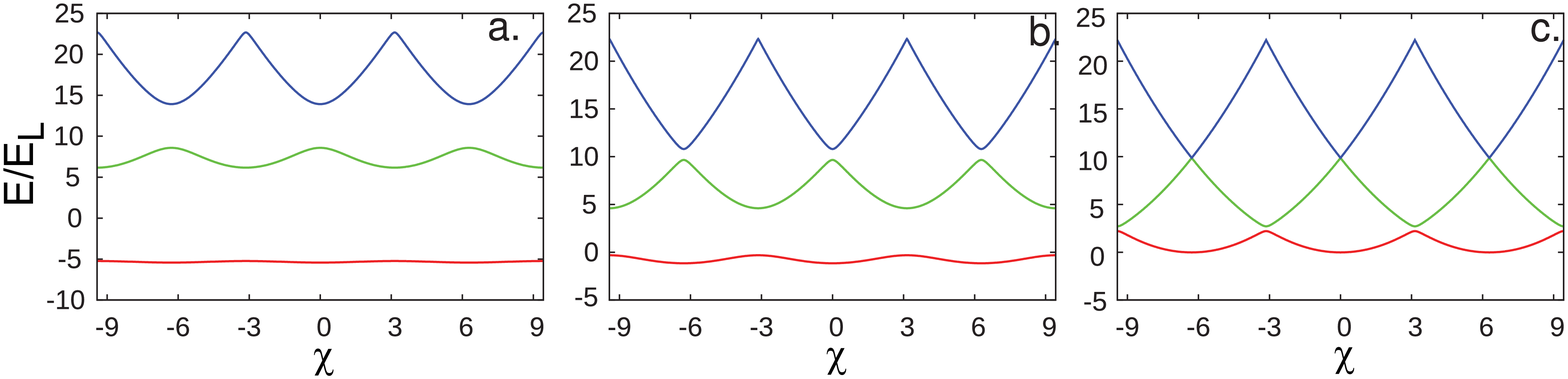}
\caption{QPS-box: Small impedance regime - Bloch states: The energies versus quasi-phase $\chi$. a: Strong PS:$E_S=6.25 E_L$. b: Moderate PS: $E_S=2.5 E_L$. c: Weak PS: $E_S= 0.25 E_L$. }
\label{fig:Bloch} 
\end{figure}

The properties of Bloch states are determined by competition of the energy scales $E_S$ and $E_L$.
If $E_L \gg E_S$, they form nearly parabolic subbands (see Fig. \ref{fig:Bloch}.c). The energy correction to the ground state appears in the second order in $E_S$,  $E^{(2)}_{0} \propto \frac{E_S^2}{E_L}$. )
Note that in distinction from the first-order correction, Eq. \ref{eq:PSB-corr-ground}, this is not exponentially suppressed.

In the opposite limit of $E_L \ll E_S$, we have a well-developed periodic potential in $Q$-space. The lowest Bloch subbands correspond to quantized energy levels in equivalent potential wells. Near the potential minimum, the energies of these levels are those of an effective oscillator, $E_n = \pi\sqrt{E_S E_L}(2n+1)$. Their dispersion comes about the tunneling through potential barriers of the height $\simeq E_S$ and is exponentially suppressed(see Fig. \ref{fig:Bloch}.a). For the lowest subband, where the suppression is the strongest,
$$
E = -|E_S| +\pi\sqrt{E_L E_S}+ 2\Delta_0 \cos(\chi),
$$
$\Delta_0\equiv 8(E_S^3 E_L)^{1/4}\cdot \exp\left(-\frac{8}{\pi}\sqrt{\frac{E_S}{E_L}}\right)$ being the amplitude of tunneling between the lowest-energy states in the neighboring wells.

\begin{figure}[!t]
\centering
\includegraphics[width=0.7\columnwidth]{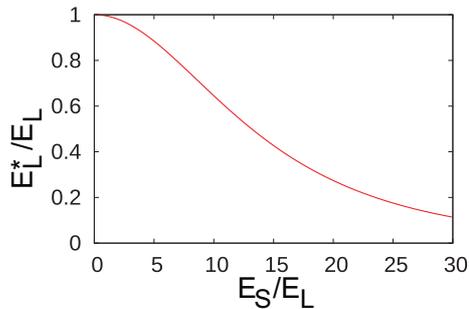}
\caption{The renormalized inductive energy $E^*_L$ versus the phase-slip amplitude $E_S$. It becomes exponentially suppressed at $E_S \gg E_L$.}
\label{fig:Bloch-curv} 
\end{figure}

Let us now resume our analysis of QPS-box and consider small but finite charging energy $E_C$. Thereby we lift the degeneracy of the potential minima that would in principle lead to charge sensitivity. Let us concentrate on the lowest energy eigenstates. Those should originate from the Bloch states with lowest energy. Their dispersion can be approximated as ${\rm const} + E_L^*/4 \chi^2$, where we expand in the quasiphase $\chi$ and introduce an effective inductive energy, $E_L^*$. The latter is given by the second derivative of the lowest energy band energy with respect to $\chi$ and represents the renormalization of inductance by the phase-slip processes. We plot this quantity versus $E_S/E_L$ in Fig. \ref{fig:Bloch-curv}. 

With this, the lowest energy states can be approximated by an effective Hamiltonian in quasi-phase representation,
\begin{equation}
H =  \frac{E_L^*}{4} \chi^2 + E_C \frac{\partial^2}{\partial \chi^2},
\label{eq:renormalized_oscillator}
\end{equation}
which is one of a harmonic oscillator with the renormalized oscillation frequency $\sqrt{E_L^*E_C}/\hbar$.
These states are not charge-sensitive in this approximation of a renormalized oscillator. We expect the charge sensitivity to set on only if the approximation breaks down.

\begin{figure}[!t]
\centering
\includegraphics[width=0.7\columnwidth]{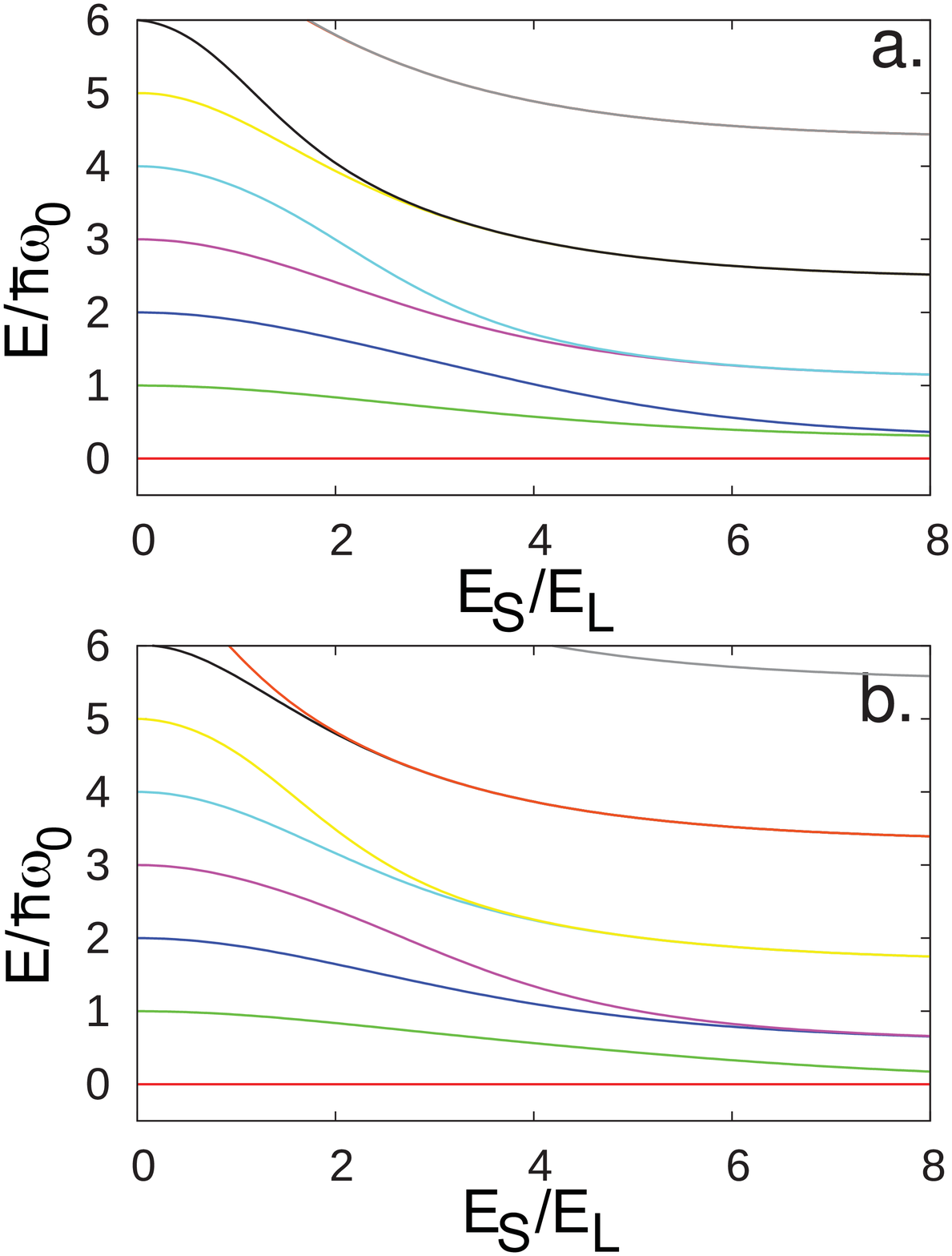}
\caption{Small impedance regime ($\gamma=1.91$): the excitation energies versus $E_S/E_L$ for two values of the induced charge of a: $q/e=0$ and b: $q/e=1$. The "sticking together" of the excitation energies at sufficiently large $E_S$  indicates emergence of the charging states.}
\label{fig:small_imp_lvls} 
\end{figure}
 
The validity of the approximation  can be estimated by comparing the spread of the ground state wave function in quasi-phase space ($\simeq (E_C/E_L^*)^{1/4}$) and $2\pi$, the scale at which we expect the dispersion of Bloch states to deviate from the quadratic law. We thus expect the approximation to break down, and the charge sensitivity to set on, when $E_C \simeq E^*_L$.  Since in the limit under consideration $E_L \gg E_C$, this can only become possible if the renormalized inductance is strongly suppressed. This requires $E_S \gg E_L$ and
yields $E^*_L \approx \Delta_0$. Comparing $\Delta_0$ and $E_C$, we obtain the estimation for $E_S$ at which the ground state becomes charge-sensitive,
$$
E_S^* \simeq E_L \left(\frac{\pi}{8}\right)^2\ln^2\left( \frac{E_C}{E_L} \sqrt{\frac{8}{\pi^3}}\right).
$$

\begin{figure}[!t]
\centering
\includegraphics[width=\columnwidth]{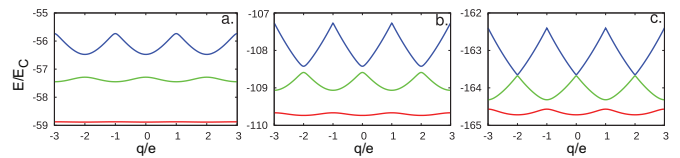}
\caption{The charge sensitivity in the small impedance regime ($\gamma = 1.91$) for a: $E_S/E_L=5.48$, b: $E_S/E_L = 8.22$ and c: $E_S/E_L = 10.96$.}
\label{fig:small_imp_qdep}
\end{figure}

In this limit, we can restrict our consideration to the states of the lowest Bloch subband assuming that the energy is small in comparison with the subband splitting, $E \ll \sqrt{E_L E_S} \gg E_C$. This is equivalent to taking into account only the lowest-energy state in each potential well. We number these states with $N$ and arrive at the effective Hamiltonian

\begin{equation}
H = -\Delta_0 \sum_{N} \left(|N\rangle \langle N+1|+|N+1\rangle \langle N|\right) + E_C(\hat{N} -q/2e)^2
\label{eq:small-imp-hopping}
\end{equation}

which appears to be the same as  the Hamiltonian Eq. \ref{eq:H_CPB_charge} of the Cooper-pair box with $E_J$ replaced with $2\Delta_0$. The energy levels follow the standard Coulomb blockade pattern provided $E_C \gg \Delta_0$. Therefore, we have proven the equivalence of CPB and QPS-box
in the limit of large $E_S$ and the emergence of well-developed Coulomb blockade in the QPS-box.

We complement this analytical consideration valid in the limit $\gamma \gg 1$ by numerical calculations at finite value $\gamma =1.91$. In Fig. \ref{fig:small_imp_lvls} we plot the energies of several excited states counted from the ground state versus $E_S/E_L$. To illustrate the charge sensitivity, the plots are made at $q/e=0$ and $q/e=1$. We see how the states evolve from equally separated oscillator levels at $E_S=0$ to charging states at $E_S $ that follow a typical Coulomb blockade degeneracy pattern (at $q/e=0$ only the excited states are doubly degenerate while at $q/e=1$ the ground state is doubly degenerate as well). We see that the higher excited states become more charge-like at smaller values of $E_S$, due to quadratic dependence of the charging energy on the state number. This is also illustrated in Fig. \ref{fig:small_imp_qdep} where we plot the energies of the ground and first two excited states versus induced charge upon the increase of $E_S$.

\begin{figure}[!h]
\centering
\includegraphics[width=0.7\columnwidth]{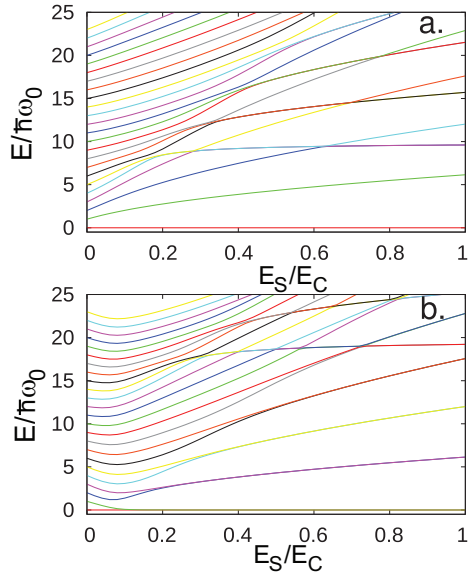}
\caption{Excitation energies in the large impedance regime ($\gamma = 0.32$) for a: $q/e=0$ and b: $q/e=1$.}
\label{fig:big-imp-lvls} 
\end{figure}

\begin{figure*}
\centering
\includegraphics[width=0.7\textwidth]{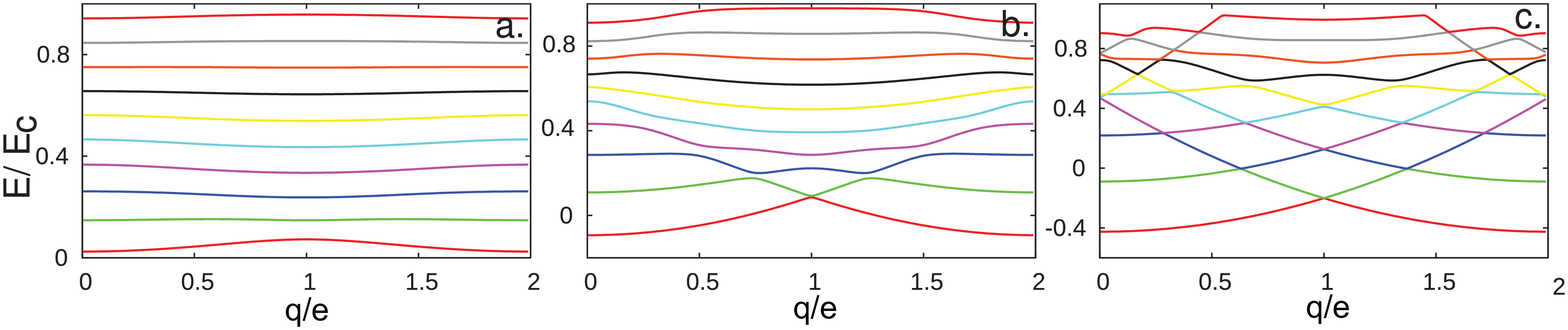}
\caption{The charge sensitivity in the large impedance regime ($\gamma = 0.32$) for a: $E_S/E_C=0.02$, b: $E_S/E_C=0.1$ and c: $E_S/E_C=0.3$.}
\label{fig:big-imp-qdep} 
\end{figure*}

\subsection{Large impedance regime} 

In this limit, $E_C \gg E_L$ and $\gamma \ll 1$.
To start with, let us disregard the inductive energy. The Hamiltonian is diagonal in charge representation and reads 
\begin{equation} \label{eq:pot_gamES}
H =  E_C \left(Q - \frac{q}{2e} \right)^2 -2 E_S \cos (2 \pi Q).
\end{equation}

The stable states are associated with the minima of $E(Q)$. The positions of minima $Q_m$ are determined from the equation 

\begin{equation}
\frac{E_C}{E_S} = -2 \pi\frac{\sin(2 \pi Q_m)}{Q_m-\frac{q}{2e}}
\end{equation}

the corresponding energies being given by

\begin{equation}\label{eq:EcQ}
E_m = E_C \left( (Q_m -\frac{q}{2e})^2+\frac{Q_m-\frac{q}{2e}}{\pi} \cot (2 \pi Q)\right)
\end{equation}

If $E_S \ll E_C$, the energy has a single minimum, and exhibits weak charge sensitivity $E(q)=-2 E_S \cos(\pi q/e)$, in agreement with Eq. \ref{eq:PSB-corr-ground}. At $q/e=1$ and critical value of $E_S=  E_C/8 \pi^2$ this minimum splits into two.
More minima emerge upon increasing $E_S$. Finally, at $E_S \gg E_C$, a large number ($\simeq \pi E_S/E_C$) of low-energy minima are pinned by the oscillatory potential to $Q_m = N $.
Their energy is contributed by the charging energy term only and reproduces the Coulomb blockade pattern
$E_N = E_C(N-q/2e)^2$. We thus conclude that the crossover to the well-developed Coulomb blockade takes place at $E_S \simeq E_C$.
The crossings of the energy levels are not avoided in this approximation. 

\begin{figure} [!t]
\centering
\includegraphics[width=0.5\columnwidth]{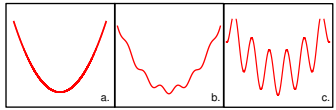}
\caption{The evolution of $E(Q)$ at $q/e=1$ upon increasing the phase-slip amplitude for a:  $E_S/E_C = 0 $, b: $E_S/E_C = 0.5 $ and c: $E_S/E_C = 10 $. The multiple potential minima formed correspond to discrete charging states.}
\label{fig:potential}
\end{figure}

Let us take into account the final value of $E_L$. This leads to quantization of energy levels around each minimum that is described 
by an effective oscillator Hamiltonian,
$$
E^*_C (Q_m - Q)^2 - \frac{E_L}{4} \frac{\partial^2}{\partial Q^{2}}
$$
where $E^*_C \equiv 2 E''(Q_m)$ is the charging energy renormalized by the phase-slips. Since $E_L$ is small, the the distance between the levels is  smaller than the  scale of the potential $E(Q)$. At $E_S \gg E_C$ we deal with almost identical potential wells of depth $4 E_S$. The oscillatory mode in each well gives rise to a system of equidistant levels separated by $2 \pi \sqrt{E_S E_L}$. Finite $E_L$ also induces tunneling between the neighboring wells that removes the degeneracy at the level crossings.

We distinguish two cases in the regime of well-developed Coulomb blockade $E_S \gg E_C$. At sufficiently small $E_S$, the effective oscillator frequency $2\pi \sqrt{E_S E_L}$ is much smaller than the Coulomb energy. In this case, the lowest-energy states are charging states (labeled by $N$) with $n$ oscillator quanta. If one neglects the exponentially small tunneling between the wells, the energies of these states are given by
$$
E(N,n) = \pi\sqrt{E_SE_L} (2n+1) + E_C(N-q/2e)^2
$$
Tunneling between the wells leads to a fascinating picture of avoided crossings  between the energy levels that differ in $N$ and $n$ (see Fig. \ref{fig:big-imp-qdep}.c)

In the opposite case $E_S \gg E^2_C/E_L$, the first excited state of the oscillator lies much higher than many charging states. The tunneling in this case is between the ground states in each potential well. The low-energy states are described by the same Hamiltonian Eq. \ref{eq:small-imp-hopping} as in the regime of small impedance. However, in distinction from the small impedance regime, $\Delta_0$ is always parametrically smaller than $E_C$.

\begin{figure}[!h]
\centering
\includegraphics[width=0.7\columnwidth]{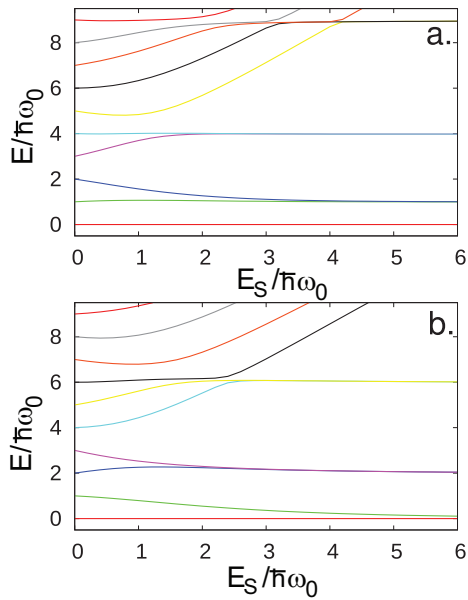}
\caption{Excitation energies in the intermediate impedance regime ($\gamma =1 $) for (a)$q/e=0$ and (b) $q/e=1$.}
\label{fig:int-imp-lvls} 
\end{figure}

\begin{figure}[!h] 
\centering
\includegraphics[width=\columnwidth]{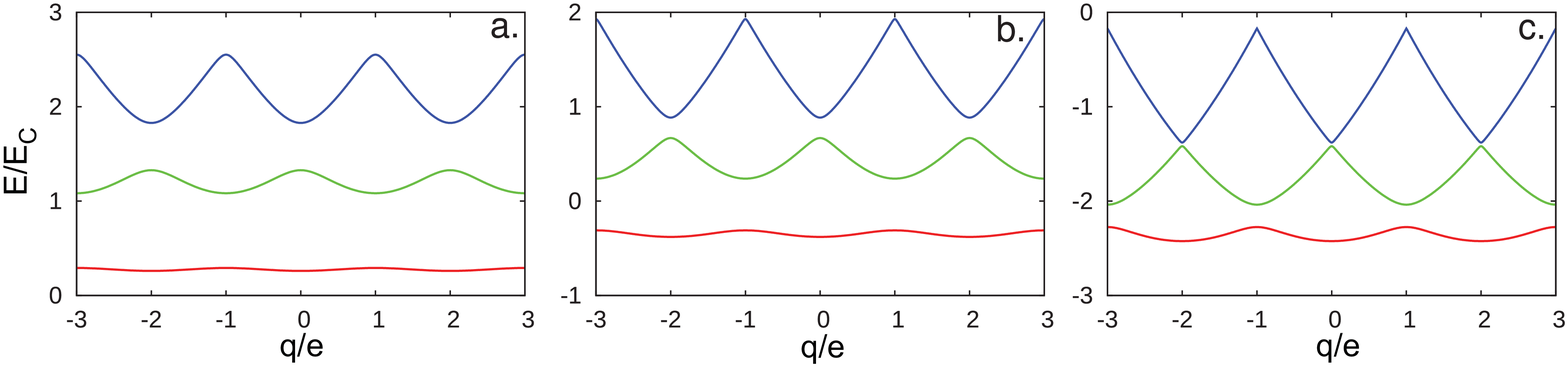}
\caption{The charge sensitivity at the intermediate impedance ($\gamma = 1$) for a: $E_S/\hbar \omega_0=1.0$, b: $E_S/\hbar \omega_0=2.0$ and c: $E_S/\hbar \omega_0=4.0$.}
\label{fig:int-imp-qdep}
\end{figure}

The above analytical analysis is confirmed by numerical calculations at $\gamma = 0.32$. In Fig. \ref{fig:big-imp-lvls} we see the evolution of the excitation energies from those of the original oscillator to those of charging states (seen as horizontal lines at sufficiently large $E_S$) or charging states with extra number of quanta of the renormalized oscillator (seen as rising lines owing to $E_S$ dependence of the oscillator frequency). The charge sensitivity is illustrated in Fig. \ref{fig:big-imp-qdep} . We plot a number of low-lying states versus $q$ at increasing values of $E_S$. We see the formation of charging and oscillator states and progressive reduction of anticrossings between those.

\subsection{Intermediate impedance regime and summary}
\label{sec:ES_gam=1}

To investigate the regime of intermediate impedance, we have numerically computed the eigenenergies  of the Hamiltonian Eq. \ref{eq:H_PSB}
at $\gamma=1$. The results for excitation energies are presented in Fig. \ref{fig:int-imp-lvls}.

As in previous regimes, we observe a crossover from equidistant levels of the original oscillator  to the charging states. The natural scale of for this crossover is neither $E_L$ nor $E_C$, rather, this is the frequency of the original oscillator $\hbar \omega_0 \equiv \sqrt{E_L E_C}$. As in the regime of large impedance, the charging states are augmented with quantized oscillations around the minimum of each potential well: excitations with $n\ne0$ are manifested as the curves rising with increasing $E_S$. In distinction from the large impedance case, the energy of these excitations slightly exceeds the energy of the lowest charging states once they are formed. We note that for small impedance the excitation energies typically decrease with increasing $E_S$ while for large impedance they typically grow. In intermediate case, we see that many excitations just keep approximately the same energy while converted from an oscillator level to a charging state.

Charge sensitivity (Fig.\ref{fig:int-imp-qdep}) shows a pattern similar to that in small impedance regime. This proves that at sufficiently large $E_S$ in all cases we reach a well-developed Coulomb blockade.

\begin{figure}[!h]
\centering
\includegraphics[width=0.7\columnwidth]{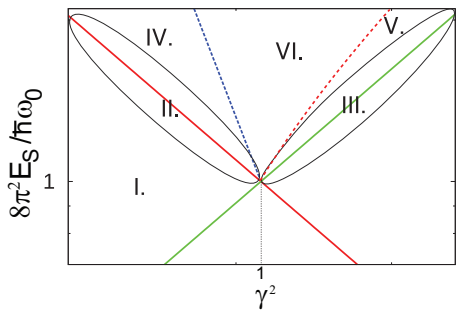}
\caption{Distinct regimes in QPS-box. The lines in his schematic log-log plot present parameter regions in the space of $E_S$ and effective impedance $1/\gamma^2$. I. Perturbative regime. II. Oscillator with renormalized capacitance. III.Oscillator with renormalized inductance.
IV. Discrete charging states accompanied by oscillator excitations. V. Direct mapping to CPB. VI. "Pure" charging states. }
\label{fig:diagram} 
\end{figure}

We summarize the regimes of QPS-box in Fig. \ref{fig:diagram}. This is a log-scale diagram with $E_S$ on the vertical axis and $\gamma^2$ on the horizontal one. Two crossing thick lines indicate $E_L \simeq E_S$ and $E_C \simeq E_S$, correspondingly. Below the lines, $E_S \ll {\rm max}(E_L,E_C)$ phase-slips can be considered perturbatively (region I). At the lines, the main effect of phase-slips is the renormalization of either capacitance (region II) or inductance (region III) of the original oscillator. Right above the line $E_S \simeq E_C$ we have a region IV were the charging states are developed and each  is accompanied by a set of closely-spaced oscillator levels. At the dashed line $E_S \simeq E^2_C/E_L$, the effective oscillator frequency becomes comparable with the charging energy. In the region V, the QPS-box is described by the CPB Hamiltonian Eq. \ref{eq:small-imp-hopping}. At the dashed line, $E_C \simeq \Delta_0$. Finally, the region VI corresponds to the standard pattern of the charging states where the degeneracy in the crossing points is lifted by the tunneling between the ground states in each potential well, $\Delta_0$ being the tunneling amplitude.

\section{QPS-transistor} \label{sec:QPS-transistor}

The physics of the QPS-transistor is determined by the same energy scales as those of the QPS-box: inductive $E_L$, charging $E_C$, and phase-slip $E_S$ energies.  The relations between the scales, as outlined in the end of the previous Section, determine the qualitative features of the QPS-transistor.
In this Section, we thus concentrate on the features of QPS-transistor that are different from QPS-box or just do not exist there. As mentioned, the levels of QPS-transistor are sensitive to flux: the feature absent in QPS-box. A symmetric QPS-transistor ($E_{S,1}=E_{S,2},E_{L,1}=E_{L,2}$) exhibits a peculiar separation of quantum variables that leads to double degeneracy of the levels at $\Phi=\pi$. We present the numerical results at intermediate impedance.

\subsection{Second-order corrections}   
As noted in Section \ref{sec:description}, the phase-slip corrections to the energy levels of the QPS-transistor are of the second order in the phase-slip amplitudes. The non-perturbed states are labeled with the phonon number $n$ and the winding number $p$. If we restrict the external phase $\Phi$ to the interval $[-\pi,\pi]$ the ground state corresponds to $|0,0\rangle$.
The correction to the ground state energy reads
\begin{equation*}
E^{(2)}_g = \sum_{n,\pm} \frac{|\langle 0,0|H_S|n,\pm1\rangle|^2}{E_0 -E_{\pm 1} - \hbar \omega_0 n},
\end{equation*}
$E_p$ being inductive energies.
To evaluate this expression, we represent it in the form of an integral over an auxiliary variable $t$,
\begin{align*}
E^{(2)}_{g} &= -\sum_{n} \int_{0}^{\infty} dt \Big[ e^{(E_{0}-E_{1} -\hbar \omega_0 n) t} \cdot |\langle 0,0|H_S|n,1\rangle|^2 \\
 &+ e^{(E_{0}-E_{-1} -\hbar \omega_0 N) t} \cdot |\langle 0,0|H_S|n,-1\rangle|^2  \Big].
\end{align*}

After this, the sum over $n$ can be taken and the integral can be evaluated.
The result reads
\begin{widetext}
\begin{align} \label{eq:2nd_order_correction_full}
E^{(2)}_{g} & =- \frac{E_{S,1}^2}
{\pi^2(E_{L,1}+E_{L,2})} e^{-\pi^2 \gamma^2 l_1^2} \left[ \frac{1}{a_+} \ _1 F_1 \left( a_+ \pi^2 \gamma^2;1+a_+ \pi^2 \gamma^2; l_1^2 \pi^2 \gamma^2 \right) +( a_+ \rightarrow a_-)\right] \notag\\
& \ \ - \frac{E_{S,2}^2}{\pi^2(E_{L,1}+E_{L,2})} e^{-\pi^2 \gamma^2 l_2^2}\cdot \left[ \frac{1}{a_+} \ _1 F_1 \left( a_+ +\pi^2 \gamma^2;1 + a_+ \pi^2 \gamma^2;l_2^2 \pi^2 \gamma^2 \right) +( a_+ \rightarrow a_-)\right]  \notag\\
& \ \ - \frac{2 E_{S,1} E_{S,2}}{\pi^2(E_{L,1}+E_{L,2})} \cos \left(\pi \frac{q}{e}\right) e^{-\pi^2 \gamma^2 \left( \frac{l_1^2}{2} + \frac{l_2^2}{2} \right)} \cdot \left[  \frac{1}{a_+} \ _1 F_1 \left( a_+ \pi^2 \gamma^2;1+a_+ \pi^2 \gamma^2; -l_1 l_2 \pi^2 \gamma^2 \right) +( a_+ \rightarrow a_-)\right]  .
\end{align}
\end{widetext}

We have also introduced the notations $l_1, l_2$ for the ratio of the inductive energies and $a_{\pm}$ for the dimensionless energy differences:
\begin{align*}
l_1 = \frac{E_{L,1}}{E_{L,1}+E_{L,2}} ; \ l_2 = \frac{E_{L,2}}{E_{L,1}+E_{L,2}} ;\\
a_\pm = -\frac{E_{0}- E_{\pm1}}{\pi^2(E_{L,1}+E_{L,2})} = l_1l_2(1\pm \Phi/\pi)
\end{align*}

The correction Eq. \ref{eq:2nd_order_correction_full} naturally separates into two parts: the "classical" one, that is contributed by the squares of the phase-slip amplitudes in two wires, and the interference one, that bears charge dependence.
Its rather complicated expression has simple asymptotes at large and small $\gamma$. 

In the limit of large impedance ($\gamma \ll 1$), the correction is determined by the contribution of the virtual state with $n=0$ and reads
\begin{align}
E^{(2)}_{g} & = - \frac{E_{S,2}^2+E_{S,1}^2 +2 E_{S,1}E_{S,2}\cos{\left( \pi \frac{q}{e}\right)}}{\pi^2(E_{L,1}+E_{L,2})}\left( \frac{1}{a_+} +\frac{1}{a_-}\right). 
\label{eq:E2_smallgam}
\end{align}
The "classical" and interference part is of the same order of magnitude.

In the opposite limit ($\gamma \gg 1$), the "classical" part remains of the same order of magnitude  while the charge-sensitive interference correction is exponentially suppressed
\begin{align}
E^{(2)}_{g} & = -  \frac{E_{S,1}^2}{\pi^2(E_{L,1}+E_{L,2})}  \left[ \frac{1 }{a_+ + l_1^2} + \frac{1 }{a_- + l_1^2} \right] \notag\\
&\ \ - \frac{E_{S,2}^2}{\pi^2(E_{L,1}+E_{L,2})}  \left[ \frac{1 }{a_+ + l_2^2\gamma^2} + \frac{1 }{a_- + l_2^2) } \right] \notag \\
&\ \ -  \frac{2E_{S,1}E_{S,2}}{\pi^2(E_{L,1}+E_{L,2})} \cos{\left( \pi \frac{q}{e}\right)} \pi \gamma \sqrt{\frac{2\pi}{l_1l_2(1-|\Phi|/\pi)}}\notag \\
&\cdot   e^{-\frac{\gamma^2}{2} [1+ 2l_1l_2 ((1-|\Phi|/\pi) (\ln(e(1-\Phi/\pi)) -1)] }\label{eq:E2_biggam}.
\end{align}
Interestingly, the exponent of this suppression varies with $\Phi$ changing between $\pi^2 \gamma^2/2$ and $\pi^2 \gamma^2(1-2\l_1\l_2)/2$.

The correction diverges near the degeneracy points $\Phi =\pm \pi$. This divergence is trivial, indicating the mixing of two crossing states
$|0,0\rangle$ and $|0,\pm 1\rangle$ by the phase-slip amplitude.
 The correction to the ground state near the divergence point reads
$$
\delta E_g = -\sqrt{\epsilon^2/4 + |M|^2} + \epsilon/2,
$$   
where $\epsilon \equiv \pi^2(E_{L,1}+E_{L,2}) (1-|\Phi|/\pi)$,  and $M$ is the matrix element of $H_{S,1}+H_{S,2}$ between the degenerate states,
\begin{equation}\label{eq:level_crossing}
M = - E_{S,1} e^{i \pi q/e} e^{-\frac{\pi^2 \gamma^2}{2} l^2_1} - E_{S,2}  e^{-\frac{\pi^2 \gamma^2}{2} l^2_2}
\end{equation}
This matrix element is exponentially suppressed in the limit of small impedance.

\begin{figure}
\centering
\includegraphics[width=0.7\columnwidth]{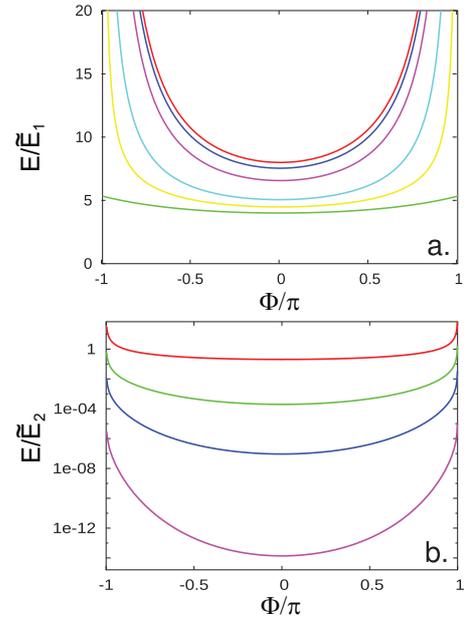}
\caption{QPS-transistor: the second-order correction to the ground-state energy (in units of $\tilde{E}_1=-E_{S,1}^2/(4(E_{L,1}+E_{L,2}))$ and respectively $\tilde{E}_2=-E_{S,1} E_{S,2}/(4(E_{L,1}+E_{L,2}))$ ). (a) "Classical" part of the correction versus external phase difference $\Phi$. From the uppermost curve downwards:  $\gamma \ll 1$ (asymptote), $\gamma=0.5$,  $\gamma=1$,  $\gamma=2$,  $\gamma=3$,  $\gamma \gg 1$ (asymptote). (b) Charge-sensitive ("interference") part of the correction versus $\Phi$. From the uppermost curve down: $\gamma \ll 1$ (asymptote), $\gamma=4$,  $\gamma=5.65$ and $\gamma=8$. Note the log-scale of the plot.}
\label{fig:QPS-asymp}
\end{figure}

We present the numerical estimations of the second-order corrections in Figs. \ref{fig:QPS-asymp}.a,b. We assume the symmetric QSP-transistor.
The "classical" correction at any value of the impedance  takes values between the two asymptotes, Eq. \ref{eq:E2_smallgam} and Eq. \ref{eq:E2_biggam}. At each value of 
impedance, the correction diverges upon approaching $\Phi = \pm \pi$.  However the divergence controlled by the matrix element Eq. \ref{eq:level_crossing} is exponentially suppressed at large $\gamma$. This is why the curves approach the non-divergent asymptote Eq. \ref{eq:E2_biggam} in the limit of small impedance. 

The charge-sensitive interference contribution is plotted in Fig. \ref{fig:QPS-asymp}.b in logarithmic scale to show the exponential dependence at large 
$\gamma$. Since the exponent depends on $\Phi$, the correction also displays exponential flux sensitivity.

\subsection{Large $E_S$}
Let us turn to the QPS-transistor in the opposite regime of large $E_S$ where the phase-slips produce the well-developed potential wells that house the charging states and oscillators. As we have seen in the case of QPS-box, this situation is achieved for sufficiently large $E_S$ for any impedance. 

We concentrate on the QPS-transistor Hamiltonian in the charge representation,
\begin{align}
H &= -\frac{E_{L,1}}{4} \frac{\partial^2}{\partial Q_1^2} - \frac{E_{L,2}}{4} \left(\frac{\partial}{\partial Q_2} + i \Phi\right)  \notag \\ 
& \ \ - 2 E_{S,1} \cos{\left(2\pi Q_1\right)} - 2 E_{S,2} \cos{\left( 2 \pi Q_2\right)} \\
&\ \ + E_C(Q_2-Q_1 -q/2e)^2\label{eq:H0-PSET-ch-rep} .
\end{align}
where the wave function is subject to the periodicity condition 
$$
\Psi(Q_1+1, Q_2+1) = \Psi(Q_1,Q_2).
$$
The phase-slip terms pinpoint the low-energy wavefunctions to the minima of the oscillating potential $Q_{1,2} = N_{1,2}$. 
Each minimum corresponds to a charging state and gives two series of oscillator levels labeled $n_{1,2}$.
The states are thus labeled as $|N_1,N_2,n_1,n_2\rangle$
and their energies  are given by
\begin{align}
E(N_{1,2},n_{1,2}) &= - |E_{S,1}| - |E_{S,2}| + E_C(N_2 -N_1 - q/2e)^2 \notag \\
& \ \ + \hbar\omega_1 n_1 + \hbar\omega_2 n_2 
\end{align}
with the oscillator frequencies being $\hbar\omega_{1,2} = 4\pi \sqrt{E_{S,1,2} E_{L,1,2}}$.

Interestingly, for a QPS-transistor we get in this limit  two oscillators instead of a single original one.    

Let us take into account the tunneling between the potential wells and do it for the lowest states $n_1,n_2=0$ only.
The tunneling amplitudes in the directions $Q_1$, $Q_2$ are $\Delta_1$, $\Delta_2 e^{i\Phi}$ respectively, the latter factor incorporates the
external phase. The expressions for $\Delta_{1,2}$ are identical to that of the QPS-box,
\begin{align*}
\Delta_1 &= 8 (E_{S,1}^3 E_{L,1})^{1/4}\cdot \exp\left(-\frac{8}{\pi}\sqrt{\frac{E_{S,1}}{E_{L,1}}}\right);\\
\Delta_2 &= 8(E_{S,2}^3 E_{L,2})^{1/4}\cdot \exp\left(-\frac{8}{\pi}\sqrt{\frac{E_{S,2}}{E_{L,2}}}\right).
\end{align*}
The tunneling part of the Hamiltonian is thus 
\begin{align}
H_{{\rm tunn}} &= - \sum_{N_1,N_2} \Big( \Delta_1 |N_1,N_2\rangle \langle N_1 +1, N_2| \notag \\ & \ \ \ \ \ \ \ \ \ + \Delta_2  e^{i\Phi} |N_1,N_2\rangle \langle N_1 , N_2+1| +h.c \Big)
\label{eq:PSET-tunneling}
\end{align}

Let us now use the periodicity condition that allows us to identify the states with $N_2-N_1=N$ and disregard higher oscillator states assuming $\hbar\omega_{1,2} \gg \Delta_{1,2}, E_C$.  With this, the Hamiltonian reduces to 
\begin{equation}
H = \sum_{N} \left(-(\Delta_1 +\Delta_2 e^{i\Phi}) |N\rangle\langle N+1| +h.c.\right) + E_C (N-q/2e)^2
\end{equation}
This is the Hamiltonian of the CPT where the Josephson energies $E_{J,1}, E_{J,2}$ are replaced with $2 \Delta_{1,2}$.
This reduction is less trivial than that in the case of QPS-box since QPS-transistor has in principle more degrees of freedom.

In the limit of $E_C \ll \Delta_{1,2}$, the QPS-transistor is thus equivalent to a double Josephson junction with the ground-state energy $-|\Delta_1 + \Delta_2 e^{i\Phi}|$ and the critical current $I_c \simeq (2e/\hbar) {\rm min}(\Delta_1, \Delta_2)$. 

This estimation should be compared with the maximum current through the wire in the ground state under the conditions of phase bias. Since the winding number in the ground state adjusts itself to the phase, the effective phase difference never exceeds $\pi$ and this current is given by
 $I_C (2e/\hbar)\pi E_{L,1}E_{L,2}/(E_{L,1}+E_{L,2})$ (this is much smaller than the critical current in a current-biased wire where large phase differences can be built). Since the tunneling amplitudes $\Delta_{1,2}$ are exponentially suppressed (Eq. \ref{eq:Delta12}) the
critical current in the wire is also suppressed exponentially in system parameters, $\ln Ic \simeq - \sqrt(E_s/E_L)$.
Notably, this suppression is more efficient than the intercepting the wire with tunnel barriers, that leads to a power-law suppression in terms of the system parameters, $I_c \simeq E^2_J/E_C$.

In the opposite limit of $E_C \gg \Delta_{1,2}$ the flux sensitivity of the device is determined by the second-order corrections 
to the ground state, that is, $\simeq \Delta^2/E_C$. We assume the ground state to correspond to $N=0$ and $-1<q/e<1$. The second-order correction
then reads
\begin{equation}
E^{(2)}_g = - \frac{2|\Delta_1+\Delta_2 e^{i\Phi}|^2 }{E_C(1-(q/2e)^2}.
\end{equation}

This diverges at $q/e = \pm 1$ where the energies of two charging states cross. Near the avoided crossing, the ground-state energy is given by 
\begin{equation}
\label{eq:anticrossing-large-ES}
E_g = E_C/4 - \sqrt{\epsilon^2 +|\Delta_1+\Delta_2 e^{i\Phi}|^2},
\end{equation}
where $\epsilon \equiv E_C (1-q/e) \ll E_C$ (for the crossing at $q/e =1$).

\begin{figure}
\centering
\includegraphics[width=\columnwidth]{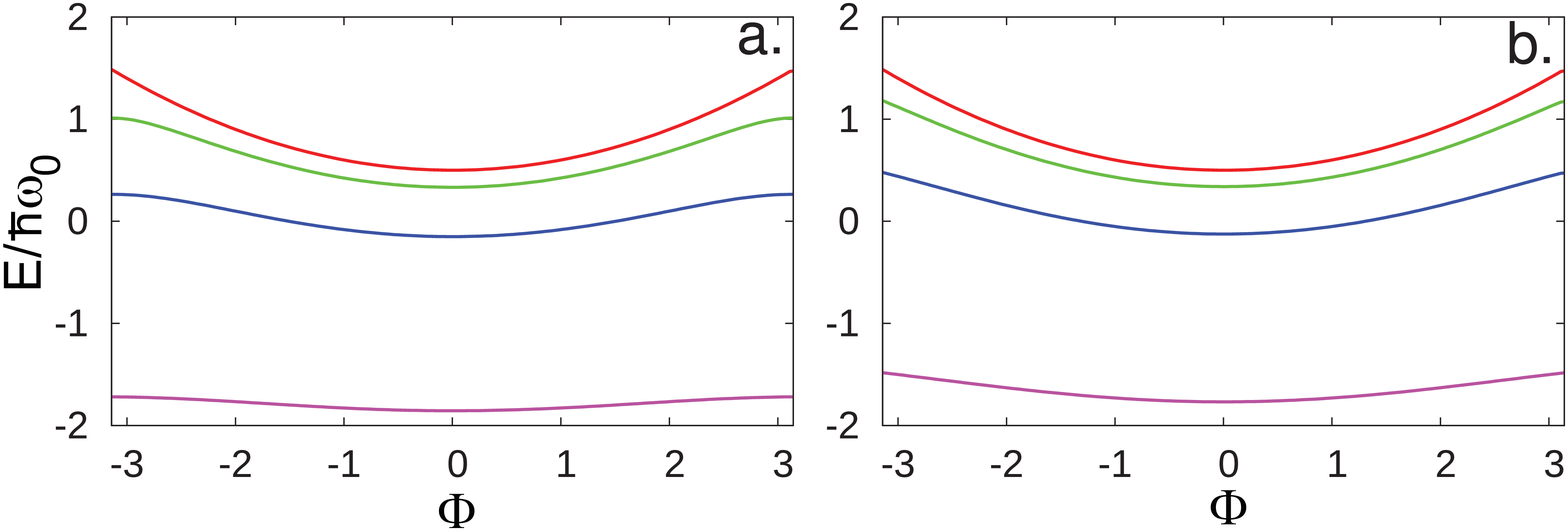}
\caption{Ground state energy of the QPS-transistor versus external phase $\Phi$ ($\gamma = 1$) for a: $q/e=0$ and b: $q/e=1$. In both panels, the phase-slip amplitudes take values: $E_S/\hbar\omega_0 = 0, 0.5, 1$ and $2$ from the uppermost curve to the lowest one.}
\label{fig:GS-flux-sens}
\end{figure}

\begin{figure}
\centering
\includegraphics[width=\columnwidth]{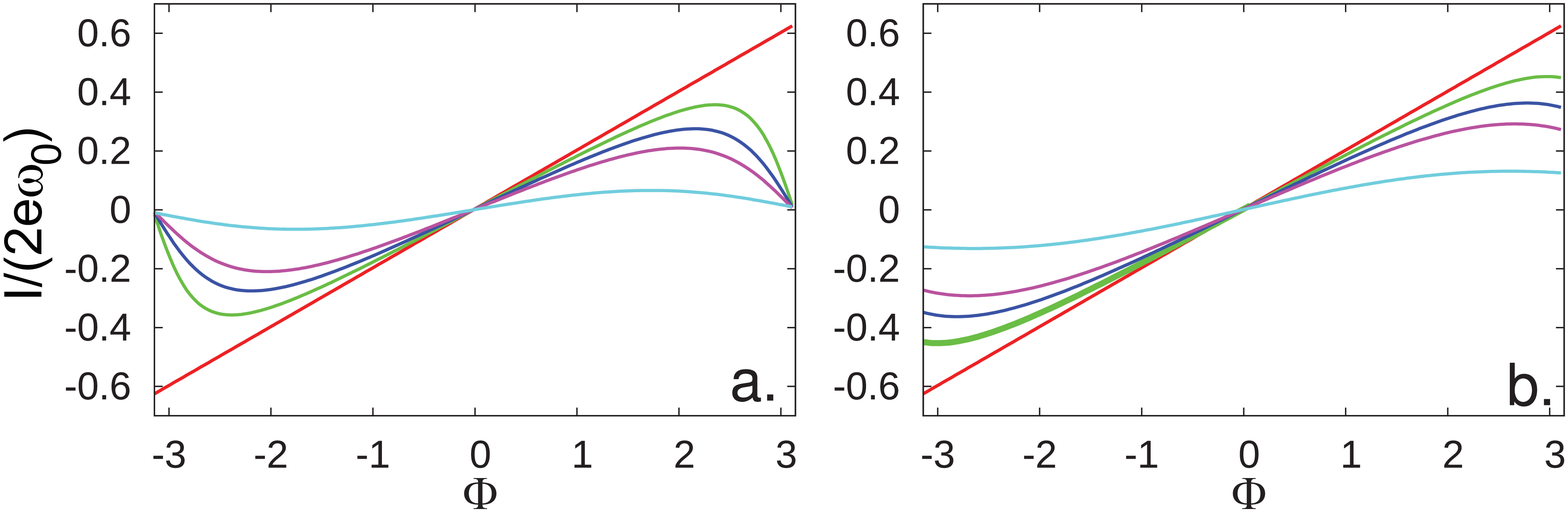}
\caption{Supercurrent in QPS-transistor versus $\Phi$ for a: $q/e=0$ and b: $q/e=1$. In both panels, the phase-slip amplitude takes values $E_S/\hbar\omega_0 = 0, 0.5,0.75, 1$ and $2$. $E_S=0$ corresponds to the straight line.}
\label{fig:current}
\end{figure}

\subsection{Degeneracies in a symmetric QPS-transistor}
The odd integer values of $q/e$ correspond to the double degeneracy of the pure charging states, 
while the pure flux states are degenerate at half-integer values of the external flux $\Phi=\pm \pi$.
 We have seen that in general the degeneracies of this type are lifted: Eq. \ref{eq:E2_smallgam} and  Eq. \ref{eq:anticrossing-large-ES}  describe avoided crossing of flux and charge states, respectively. However, the non-diagonal matrix element in both expressions vanishes if the degeneracies were to occur simultaneously, that is, at $\Phi=\pm \pi$ and $q/e=\pm 1$, and the QPS-transistor is completely symmetric $(E_{S,1}=E_{S,2},E_{L,1}=E_{L,2})$. The double degeneracy thus persists in this point.
 
Since both expressions are perturbative, one could think that the degeneracy lifting is just not visible being governed by the  next-order perturbation terms. However, this is not the case: the double degeneracy is preserved by a specific symmetry of the QPS-transistor Hamiltonian that takes place at odd values of $q/e$ and is not obvious from the forms of the Hamiltonian that we have used so far.

To reveal the symmetry, let us rewrite the Hamiltonian of the QPS-transistor in the basis of unperturbed wavefunctions $|n,p\rangle$, $n$ being the number of quanta of the original oscillator and $p$ being the winding number across the wires.
Actually, this is the representation we have made use of in numerical simulations. It reads
\begin{align}
\label{eq:Hamiltonian_in_oscillator_basis}
H &= \sum_{n,p} (E_p + \hbar \omega_0 n) |n,p\rangle \langle n,p| \notag \\
& + \sum_{n'} \left( P_{n,n'}|n,p\rangle \langle n',p+1| + h.c.\right),
\end{align}
$E_p = (2\pi p +\Phi)^2 E_{L,1} E_{L,2}/4(E_{L,1} + E_{L,2})$ being the inductive energies of the states with different winding numbers. The matrix elements of the phase-slip terms read
\begin{align}
\label{eq:matrix_elements_QPS_SET}
P_{n,n'} &= -E_{S,1} e^{i\pi q/e} \int d \phi \ \psi_n(\phi) \psi_{n'}(\phi + 2\pi l_1 ) \notag \\
& \ \ \ \ \ \ - E_{S,2} \cdot \int d \phi \ \psi_n(\phi) \psi_{n'}(\phi - 2\pi l_2 ),
\end{align}

$\psi_n(\phi)$ being the oscillator wave functions in the phase representation. It is evident from this expression that for a symmetric QPS-transistor the matrix elements obey a specific selection rule: they are non-zero only between the oscillator states of different parity, $n+n'$ should be odd. Therefore, the whole Hilbert space is separated in two disconnected blocks. In one block("+"), $n$ is even for even $p$ and odd for odd $p$, while in another block ("-") the situation is reversed: $n$ is odd for even $p$ and even for odd $p$. 

\begin{figure}\label{fig:2Dlattice}
\centering
\includegraphics[width= 0.6\columnwidth]{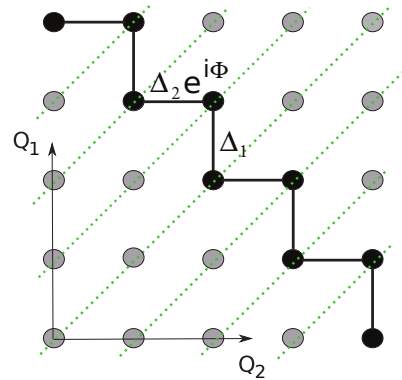}
\caption{QPS-transistor in the limit of large $E_S$. The positions of the potential minima in extended charge space $(Q_1,Q_2)$. The dotted lines connect equivalent minima. representing the set of points which are the same in this space. Dark dots give unique minima. Tunneling amplitudes through potential barriers (Hamiltonian Eq. \ref{eq:PSET-tunneling}) are indicated by Greek letters.}
\end{figure}

The energies in each block are $4\pi$-periodic functions of the external phase. Since $2\pi$ shift of the phase shifts $p$ by $1$ and therefore switches between even and odd $p$, these energies must satisfy $E^+(\phi) = E^{-}(2\pi + \phi)$. This implies the double degeneracy at all half-integer values of the external flux provided $q/e$ is odd-integer.

In fact, the even-integer values of $q/e$ are also specific for the symmetric QPS-transistor. As one can see from Eq. \ref{eq:matrix_elements_QPS_SET}, at these values the matrix elements are only non-zero between $n$ of the same parity. The Hilbert space separates into two blocks comprising respectively odd and even $n$. The energies remain $2\pi$ periodic, so this separation does not imply extra degeneracies at half-integer values of the external flux. However, the "random" degeneracies due to the crossing of the levels of different blocks at some non-specific values of $\phi$ do occur. 

Although these peculiar separations hold for a symmetric QPS-transistor only, they may become important in the course of quantum manipulation of the states of QPS-transistor. It is not excluded that the values of $E_S$ in a QPS-transistor may be tuned by some extra gates, this would be a fully symmetric QPS-transistor practical. This is why we have discussed it and give numerical results for the symmetric QPS-transistor configuration.

\begin{figure}[!t]
\centering
\includegraphics[width=\columnwidth]{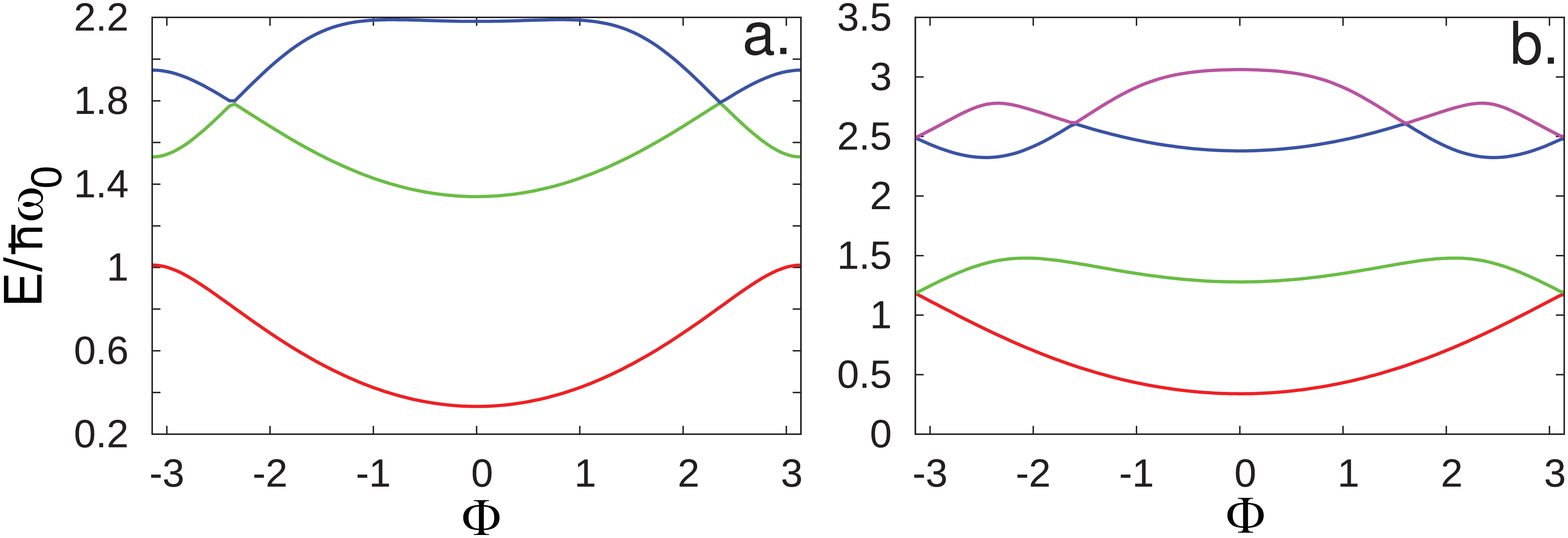}
\caption{Flux dependence of the lowest three energy levels for a "symmetric" QPS-transistor ($E_{L,1} = E_{L,2}; E_{S,1} = E_{S,2}$), for $\gamma=1$, $E_S/\hbar \omega_0 = 0.5$ and a: $q/e=0$ and b: $q/e=1$. Note the degeneracies at half-integer flux for $q/e=1$.}
\label{fig:enbands}
\end{figure}

\subsection{Flux sensitivity}
We illustrate the flux sensitivity of QPS-transistor at different values of phase-slip amplitude $E_S=E_{S,1}=E_{S,2}$. We restrict these examples to the case of intermediate impedance $\gamma=1$. 

In Fig. \ref{fig:GS-flux-sens}, we give the dependence of the ground state energies on the external phase $\Phi$ for weak, moderate and large phase-slip amplitudes. For vanishing $E_S$, the dependencies are parabolic. For $q/e=0$ (Fig. \ref{fig:GS-flux-sens}.a), we see the rounding of parabolas at $\Phi=\pm \pi$ so the curves approach the ${\rm cos}$ shape upon increasing $E_S$. Also, the flux sensitivity defined as the energy difference between $\Phi=\pi$ and $\Phi=0$, quickly decreases  upon increasing $E_S$. For $q/e=1$, the curves remain parabola-like, this is due to special degeneracy described in the previous subsection. Nevertheless the flux sensitivity decreases upon increasing $E_S$, although not as fast as in $q/e=0$ case.

The energies of the three states with lowest energy are plotted in Fig \ref{fig:enbands}. at $E_S=0.5\hbar\omega_0$. At $q/e=0$, this exemplifies the degeneracy lifting and avoided level-crossing at half-integer flux. In contrast to this, we see unavoided crossing of the first and second excited state at $\Phi \approx \pm 2.3$, this indicates opposite parity of the states. At $q/e=1$, we see the double degeneracy at $\Phi = \pm \pi$.

\begin{figure}[!t]
\centering
\includegraphics[width=0.6\columnwidth]{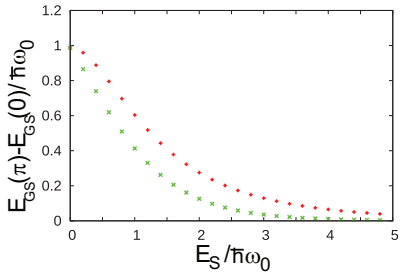}
\caption{Flux sensitivity versus phase-slip amplitude for two values of charge ($\gamma = 1$). The crosses: $q/e=0$. X-symbols: $q=1$.}
\label{fig:flux-sensit}
\end{figure}

Fig. \ref{fig:flux-sensit} gives the flux sensitivity versus $E_S$ for $q/e=0$ and $q/e=1$. They coincide at $E_S=0$ indicating no charge sensitivity. Most interesting region is that of moderate $E_S$ where the flux sensitivity is charge sensitive but still substantial. In agreement with the considerations in Subsection B, the flux sensitivity drops exponentially at large $E_S$.
Although this takes place for both values of the external charge, the flux sensitivity drops much slower for $q/e=1$ and eventually 
exceeds by far the flux sensitivity at $q/e=0$. Indeed, the same considerations show that the phase-dependent energy is $\simeq \Delta$ for $q/e=1$ and $\simeq \Delta^2$ for $q/e=0$.

\section{Dual devices} \label{sec:dual}
As discussed in \cite{Mooij2006}, there is a duality between phase-slip junctions and Josephson junctions. Each device containing phase-slip junctions has an analogue where phase-slips are replaced with Cooper-pair tunneling events, phase is replaced with charge. Upon this dual transformation, the devices have identical quantum dynamics. In this Section, we will find and shortly discuss the dual analogues of the QPS-box and QPS-transistor.

To avoid any misunderstanding: the analogies between CPB and QPS-box, and  between CPT and QPS-transistor that we have thoroughly discussed in this article are not related to the duality under consideration. Dual devices are rather different.

The duality is based on the canonical transformation $(\hat{Q}, \hat{\phi}) \rightarrow (- \hat{\phi}/2 \pi, 2 \pi \hat{Q})$ that preserves the commutation relation: $[\hat{Q}, \hat{\phi}]=-i$. With this, we can readily establish the Hamiltonians of the dual devices.
The Hamiltonian dual to that of QPS-box Eq. \ref{eq:H_PSB} reads

\begin{equation}\label{eq:H-PSB-dual}
H_{{\rm QPS-box}}^{{\rm dual}} = \frac{E_L'}{4} (\hat{\phi}- \Phi)^2 + E_C'\hat{Q}^2  - E_J \cos\hat{\phi}.
\end{equation}

It is a Josephson Hamiltonian with the parameters related to the initial QPS values as follows:
\begin{equation*}
E_J \rightarrow 2 E_S; \quad E_L' \rightarrow \frac{E_C}{\pi^2}; \quad E_C' \rightarrow \pi^2 E_L; \Phi \rightarrow \pi q/2e
\end{equation*}
In electrical terms, this is a Josephson junction in series with an LC-oscillator. The flux sensitivity of this device is obtained from the charge sensitivity of the QPS-box, $E(\Phi) = E_{{\rm QPS-box}}(q=e\Phi/\pi)$.

For the QPS-transistor the same transformation leads to the following dual Hamiltonian:
\begin{align*} 
H_{{\rm QPS-transistor}}^{{\rm dual}} &= E'_L (\hat{\phi}_2 - \hat{\phi}_1 - \Phi ) ^2 \\
&+ E'_{C,1} \hat{Q}_1^2 - E_{J,1} \cos(\hat{\phi}_1) \\
&+ (1) \rightarrow (2). 
\label{eq:H_PSB-dual}
\end{align*}
where the constraint $Q_1 +Q_2 = q/2e + p$ is imposed, $p$ being an integer number. Physically, this is an integer number of of Cooper pairs accumulated in the nodes encircled in Fig. \ref{fig:PST-dual}. We note that in the limit of vanishing $E_J$
this combination of nodes is isolated from the leads and thus can sustain integer charge. To keep consistent notations, the gate capacitance in Fig. \ref{fig:PST-dual} should be assumed vanishingly small, $C_g \ll C_{1,2}$, whereas $C_g V_g =q$.

The regime of vanishing phase-slips in original devices correspond to the regime of vanishing Josephson couplings for dual ones.
In this regime, the devices are LC-oscillators. The dual of QPS-transistor possesses  an extra degree of freedom: the number of integer charges accumulated in the isolated island. The regime of well-developed Coulomb blockade in the QPS devices corresponds to classical limit of the dual Josephson circuits, where the winding numbers in the superconducting inductors are dual analogues of discrete charges.

\begin{figure}[!t]
\centering
\includegraphics[width=0.6 \columnwidth]{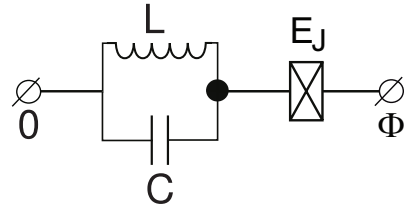}
\caption{The Josephson circuit dual to QPS-box.}
\label{fig:PSB-dual}
\end{figure}

\begin{figure}[!t]
\centering
\includegraphics[width=0.6 \columnwidth]{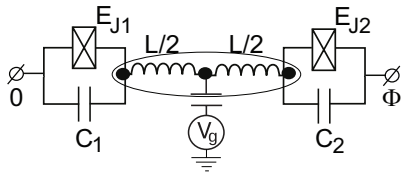}
\caption{The Josephson circuit dual to QPS-transistor.}
\label{fig:PST-dual}
\end{figure}

\section{Conclusions} \label{sec:conclusions}
In this paper we have proposed and discussed two superconducting devices made of superconducting wires subject to coherent quantum phase-slips:
QPS-box and QPS-transistor. 
Our main goal was to demonstrate the charge sensitivity, this would be unambiguous experimental signature of Coulomb-blockade behavior.
The experimental realization of our proposed devices is achievable with the state-or-the-art technology. 

We have shown that the charge sensitivity appears already for small phase-slip amplitudes as a perturbative correction to the ground-state energy. This correction is of the first order in case of QPS-box and of the second order in case of QPS-transistor. In both cases, the charge-sensitive part of the perturbative correction is exponentially suppressed in the limit of low impedance. In contrast to the QPS-box, the QPS-transistor exhibits both flux and charge sensitivity that makes it potentially useful for measurements.

However, if the phase-slip amplitude becomes of the order of either charging energy (large impedance regime) or inductive energy (small impedance regime), both devices show discrete charging states that follow a common Coulomb blockade pattern of "crossing parabolas".
The crossover to Coulomb blockade occurs differently in the limits of large and small impedance. We have revealed and investigated  both analytically and numerically six distinct parameter regions. For QPS-transistor, we have analyzed the flux sensitivity (superconducting current) as well as combined flux-charge sensitivity . For a symmetric QPS-transistor we have found a variable separation at specific values of induced charge that leads to double degeneracy of the states at $q/e=1$ and half-integer external flux. We have calculated the superconducting current through QPS-transistor to show the non-triviality of the device. We briefly discuss the Josephson-based devices that are dual to QPS-box and QPS-transistor.

\acknowledgments
The authors are indebted to J.E. Mooij who has brought the problem  to our attention and provided motivation and support to the research presented. We acknowledge fruitful discussions with K. Yu. Arutyunov, A. D. Zaikin, O. Astafiev, Yu. A. Pashkin and C. Wilson.
This work is part of the research program of the Foundation for Fundamental Research on Matter (FOM), which is part of the Netherlands Organization for Scientific Research (NWO).

\appendix
\section{Microscopic foundation}
In this Appendix, we sketch the microscopic reasoning that justifies the Hamiltonians (\ref{eq:H_PSB}),(\ref{eq:H0-PSET}). Most of this reasoning can not be regarded as the resultive part of the article. Rather, this Appendix mainly summarizes the results of Refs. \cite{review,Buchler2004,MooijHarmans,Mooij2006} perhaps with more comprehensive notations.
We include this summary for the sake of completeness, and to respond to the requests of our colleagues.

The reasoning proceeds in two steps. At the first step, one recognizes how the phase-slip amplitude emerges from instantons of an effective field theory describing the quantum fluctuations of the superconducting order parameter. The outcome is a 1d sine-Gordon-type model of the superconducting wire where the phase-slip amplitude $per\ unit\ length$ enters as a parameter. 
At the second step, one replaces this 1d model with a zero-dimensional one. This is justified if the internal degrees of freedom of the wire can be efficiently disregarded or incorporated into the parameters of the zero-dimensional model. We discuss the requirements for this. The outcome
is a Hamiltonian combining inductive energy of the wire and total phase-slip amplitude.

One can start with a Hamiltonian of electrons in the metal wire subject to Coulomb and phonon-mediated interaction. Hubbard-Statonovitch transform in imaginary-time \cite{AS} represents these two interactions in terms of the (quantum) fluctuations of two fields: $V({\bf r},\tau)$, time-dependent voltage, and $\Delta({\bf r},\tau)$, complex superconducting order parameter. Electro-neutrality condition enforces Josephson relation $2eV = \hbar \dot{\phi}$ in each point and time moment, $\phi$ being the phase of the order parameter \cite{AS}. After that, one can integrate out the electron degrees of freedom and obtain the effective action for $\Delta({\bf r}, t)$\cite{review}. If the lateral dimensions of the wire are smaller than the superconducting coherence length $\xi$, it suffices to consider only the realizations of the order parameter that are constant along the wire cross-section. Those can be parametrized with a 1d field $\Delta(x,\tau)$. The action consists of two terms,\cite{review}
$$
{\cal S} = {\cal S}_{{\rm sup}} + \left(\frac{\hbar}{2e}\right)^2\frac{\bar{C}}{2}  \int dx \frac{d\tau}{\hbar}  \dot{\phi}^2(x,\tau)
$$
Here, the first term incorporates everything related to superconductivity while the second term represents the  charging energy of the wire, $\bar{C}$ being the wire self-capacitance per unit length. 

The superconducting part of the action is non-local in both coordinate (at the scale $\simeq \xi$) and imaginary time (at the scales $\hbar/\Delta_0$, $\Delta_0$ being the saddle-point value for the superconducting gap). 
The degenerate topologically trivial saddle-point solutions correspond to constant modulus of order parameter and are parametrized by the phase,$\Delta(x,\tau) =\Delta_0 e^{i\phi}$. 
This suggests the importance of Goldstone modes that are long-wave topologically trivial fluctuations  of the phase. The effective "hydrodynamic"\cite{review} action for these fluctuations reads
\begin{equation}
{\cal S}_h = \left(\frac{\hbar}{2e}\right)^2 \int dx {d\tau} \left( \frac{(\phi')^2}{2 \bar{L}}\right) + \left(\frac{\bar{C}}{2} \dot{\phi}^2\right)
\label{eq:lw-action}
\end{equation}
$\bar L$ being the wire inductance per unit length that is mostly contributed by kinetic inductance of the superconducting material. This action describes the propagation of the electromagnetic waves along the wire with velocity $c_0 = 1/\sqrt{\bar{L}\bar{C}}$.

The full action also permits topologically non-trivial saddle-point configurations, instantons, those correspond to phase slips.
A single phase-slip configuration has a core at a point $(x_0,t_0)$ of a size $\xi, \hbar/\Delta_0$, this size corresponding to the non-locality of the action. Far from the core, the modulus of the order parameter equals $\Delta_0$, while the phase is given by
$$
\phi(x,\tau) = -\arctan\left( \frac{x-x_0}{c_0(\tau-\tau_0)}\right) 
$$
There is a significant extra action associated with a phase slip,
${\cal S}_{QPC}$. This action contains "hydrodynamic" part that comes from the long-wavelength action (Eq. \ref{eq:lw-action}) and "core" part. With logarithmic accuracy, the hydrodynamic part is given by
$$
{\cal S}_{h,QPS} = \frac{\pi\hbar^2}{4 e^2} \sqrt{\frac{\bar{C}}{\bar{L}}} \ln\left(\frac{X_{l.c.}}{X_{u.c.}}\right)
$$ 
where $X_{l.c.}$ is the lower cutoff (of the order of system size) and $X_{u.c.}$ is the upper cutoff (of the order of core size).
The core part of the action is readily estimated as
$$
{\cal S}_{c,QPS} = \frac{\alpha}{G_Q \bar{R} \xi}
$$ 
where $\bar{R}$ is the normal-state wire resistance per unit length, and $\alpha$ is a dimensionless coefficient which is not known so far. 
The evaluation of the partition function of the system in the vicinity of the phase-slip saddle-point solution gives the QPS amplitude per unit length, $\bar{E}_S \propto e^{-{\cal S}_{QPC}/\hbar}$. The total amplitude
has been estimated in \cite{review}, where it has been called "rate" $\gamma_{QPS}$. We should perhaps stress that by no means such $\gamma_{QPS}$ gives an estimate of a transition rate of any event involving quantum-phase slips, nor this was indented by the authors of $\cite{review}$. The transition rate in any case should be proportional to $\gamma^2_{QPS}$, the square of the transition amplitude. Unfortunately, the fact that the coefficient
$\alpha$ is not known makes the theory quite useless for predictions of any concrete values of $\bar{E}_S$: The exponential dependence on $\alpha$ would result in uncertainties of many orders of magnitude. The core part of the action has been found for a ballistic single-channel wire,\cite{vinokur} but this hardly helps in any realistic situation.

Further analysis presented in \cite{review,debat} concerns the instanton configurations with multiple phase-slips.
The problem can be mapped onto a 2d gas of charged particles with logarithmic interaction.    
For our purposes, it is convenient to rewrite it as an equivalent sine-Gordon model. 
A convenient variable for this is the charge $2e q(x,\tau)$ {\it passed} thought the point $x$
of the wire (the charge density $accumulated$ in the wire is then $-2e q'$ while the current 
$I=2e \dot{q}$. 
The action in terms of this variable takes the following form
\begin{eqnarray}
{\cal S} = \int dx d\tau \left(4 e^2\left( \frac{\bar{L} (\dot{q})^2}{2} +\frac{(q')^2}{2 \bar{C}}\right) \right. \nonumber\\
\left. - \bar{E}_{S} \cos(2\pi q(x,t))\right)
\label{sinGordon}
\end{eqnarray}
with two first terms giving the hydrodynamic part of the action, that is, (\ref{eq:lw-action}).
The quantity $\bar{E}_S$ corresponds to "bare" fugacity of the phase-slips, that is, contains only the "core" contribution of the action, $\bar{E}_S \propto e^{-{\cal S}_{c,QPS}/\hbar}$. The uncertainty in defining the core is incorporated into upper cutoff of the sine-Gordon model.

From a formal point of view, it seems natural to consider an infinite wire. In this case, it is very well known that the system exhibits a quantum Schmid transition (sometimes termed BKT, from analogy with associated 2d classical system) at arbitrary $\bar{E}_{S}$ and at a critical value of the wave impedance
$$
\sqrt{\frac{\bar{L}}{\bar{C}}} = \frac{\hbar}{8 e^2}.
$$ 
In the context of the wires, it is sometimes called the superconducting-insulating transition. Perhaps one should be more cautious about this, since the essence of the transition is the distinct behavior of the renormalized inductance $L$ of the wire as function of the wire length $l$, $L(l)$. The renormalization by the phase-slip always enhances the inductance, the enhancement changes from power-law $L \propto l^{\beta}$ with $\beta < 1$ to exponential dependence $L \propto e^{l/l_c}$ at the point of the transition.  Guided by personal taste, one may call the wire insulating (zero inductance per unit length in the limit $l \to \infty$) or superconducting (finite inductance at any finite $l$).

More insight into the wire properties in the insulating regime can be obtained when considering the saddle-point solutions of the sine-Gordon model. The trivial solution corresponds to $q=0$ so that no charge can pass the wire. Besides, there are sine-Gordon kinks. In this case, they correspond to the charged soliton-like excitations. The minimum energy of the soliton, $E_{{\rm sol}}\simeq e \sqrt{\bar{E}_S/\bar{C}}$, is the gap for charged excitation. The soliton is spread over the length $L_{{\rm sol}} = e/\sqrt{\bar{E}_S\bar{C}}$. The inverse inductance,that is, the supercurrent in the wire, is due to tunneling of the solitons over a potential barrier of the length $l$ and the height $E_{{\rm sol}}$. Therefore, the wire in the insulating regime is a kind of a semiconductor, a uniform barrier for charges to tunnel through. The ground state is homogeneous, one cannot say that the wire is cut into small pieces separated by tunnel barriers.   

It is not clear at the moment to which extent the 1d model can be useful to describe experimental results even at qualitative level, this being a subject of ongoing research and debate. The realistic wires are not only finite: 
they are rather short. The inverse time of electricity propagation across the wire,
$c_0/l$, can easily exceed the typical superconducting energy scale $\Delta_0/\hbar$. 
In this case, the upper cutoff is smaller than the lower cutoff, this renders hydrodynamic corrections irrelevant. 
Even if this is not the case, it has been shown \cite{Buchler2004} that for a finite wire embedded into an external circuit the Schmid transition should be governed by the external impedance of the circuit rather than the wave impedance of the wire. 
There were recent attempts \cite{debat} to modify the traditional renormalization schemes to take into account finite $l$ and possible normal excitations not captured by the hydrodynamic action. More experiments and more detailed comparison of experiment and theory are required to resolve the issue.  
 
A way to avoid these complicated issues is to think of a phase-slip superconducting wire in terms of a zero-dimensional model, that is, in terms of a circuit-theory element. For a closed wire, such model has been proposed in \cite{MooijHarmans}. The model can be derived from the sine-Gordon model (\ref{sinGordon}) by setting $q={\rm const}(x)$.        
The result is most convenient to present in the Hamiltonian form
$$
\hat{H} = E_L \hat{\phi}^{2} - E_S \cos(2\pi\hat{q}) 
$$
$E_L$ being inductive energy of the wire, $E_S = \bar{E}_S l$. The operators of phase drop across the wire,
$\hat{\phi}$ and charge passed $\hat{q}$ satisfy canonical commutation relation
$[\hat{q},\hat{\phi}]=i$. This zero-dimensional model has been extended in \cite{Mooij2006} to the case of the wire embedded into an arbitrary external circuit and its exact duality with Josephson-junction circuits has been demonstrated. 
 
The zero-dimensional model obviously does not describe the internal excitations of the wire such as standing electromagnetic waves in the absence of the phase slips or charge solitons arising from sine-Gordon model in the limit of sufficiently developed phase-slips, although the renormalization of the parameters $E_L$ and $E_S$ by these excitations can be taken into account  on phenomenological level, that is, just by taking the measurable values of these parameters rather than the "bare" ones.  One can argue that these excitations either have energies exceeding the energy scales of zero-dimensional model $E_S, E_L \gg \hbar c_0/l, E_{{\rm sol}}$ or just do not fit into the wire ($L_{{\rm sol}} \gg l$).

A convenient formal way to assure irrelevance of internal excitations is to take the limit of
vanishing self-capacitance of the wire, $\bar{C} \to 0$. In this limit, all of the above requirements are fulfilled, since the energy scale of the standing waves, soliton energy and length become infinite. The charge  cannot be accumulated in the wire, $q'(x)=0$, and the model is essentially zero-dimensional.

\end{document}